\documentclass[12pt,psf,epsf]{JHEP3}
\usepackage[centertags]{amsmath}
\usepackage{amssymb}
\usepackage{graphicx}
\usepackage{epsfig}
\usepackage{ulem}

\newcommand{\be}{\begin{equation}}
\newcommand{\ee}{\end{equation}}
\newcommand{\bea}{\begin{eqnarray}}
\newcommand{\eea}{\end{eqnarray}}

\title{Holographic phase diagram of quark-gluon plasma formed in heavy-ions collisions}
\author{I.~Ya.~Aref'eva,$^1$ A.~A.~Bagrov,$^{1,2}$ and E.~O.~Pozdeeva$^3$\\
$^1${\it Steklov Mathematical Institute, Gubkin str.~8, 119991,
Moscow,
Russia.}\\
$^2${\it Instituut Lorentz, Leiden University, Niels Bohrweg 2,
2300 RA Leiden, Netherlands}\\
$^3${\it Skobeltsyn Institute of Nuclear Physics, Lomonosov Moscow
State University,
 Leninskie gory, GSP-1, 119991, Moscow, Russia}\\
{ E-mail: arefeva@mi.ras.ru, bagrov@lorentz.leidenuniv.nl,
pozdeeva@www-hep.sinp.msu.ru}}

\abstract{The phase diagram of quark gluon
plasma (QGP) formed at a very early stage  just after the heavy
ion collision is obtained by using a holographic dual  model
for the heavy ion collision. In this dual model colliding ions are
described by the charged shock gravitational waves.
Points on the phase diagram correspond to the
QGP or hadronic matter with given temperatures and chemical
potentials. The phase of QGP in dual terms is related to
the case when the collision of shock waves leads to formation of
trapped surface. Hadronic matter and other confined states
correspond to the absence of trapped surface after collision.

A multiplicity of
the ion collision process is estimated in the dual
language as an area of the trapped surface. We show that a non-zero
chemical potential reduces the multiplicity. To plot
 the phase diagram we use two different dual
models of colliding ions, the pointlike
and the wall shock waves, and find
  qualitative  agreement of the results.
}
 \keywords{AdS/CFT,  holography, quark gluon plasma, black holes, trapped surface, shock
waves}

\begin{document}
\newpage
\section{Introduction}
For the last decade, since the publication of fascinating papers
\cite{Malda,GKP,Wit}, it was realized that supersymmetric and
non-supersymmetric theories in the strong coupling limit in
principle could be pretty close in their properties
\cite{Policastro:2001yc}. The AdS/CFT correspondence, which appeared
as a formal duality between the ${\cal N}=4$ super Yang-Mills theory and
a quantum gravity in $AdS$ background, has become powerful tool for
studying various properties of real physical systems in the strong
coupling \cite{Mateous-and-Co}.

Important branch of these investigations is the analysis of the
Quark Gluon Plasma (QGP) from the point of view of
$AdS$-holography, see for example, review \cite{HoQGP-review}.
These applications of the AdS/CFT correspondence to strongly coupled
QGP have been mostly related  to equilibrium
properties of the plasma, or to its' kinetics/hydrodynamics
 near the equilibrium.

A particular application
of AdS/CFT to the strongly coupled QGP, is the analysis of thermalization of
matter and early entropy production instantly after the collision
of relativistic heavy ions. RHIC experiments have shown that a QGP
forms at a very early stage  just after the heavy ion collision,
i.e. a rapid thermalization occurs, and QGP produced in RHIC is
believed to be strongly coupled  as evidenced by its rapid
equilibration. Strong collective flows well reproduced by
hydrodynamics, and strong jet quenching
\cite{ES1-QGP,ES2-QGP,Heinz}. This obviously requires  a
calculation of the strongly coupled field theory in non-equilibrium
process.

Not long ago Gubser, Yarom and
 Pufu \cite{Gubser} have proposed the gravitational shock wave in AdS$_5$ as a
possible holographic dual for the heavy ion and have related the
area of the trapped surface formed in a collision of such waves to
the entropy of matter formed after collision of heavy ions. Early
papers where has been mentioned an analogy between colliding heavy
ions and colliding gravitational shock waves in anti-de Sitter space
include \cite{early-G1}-\cite{early-G5}. This AdS-holographic model
has been also used to find the stress-energy tensor of the QGP
formed by ion collision. In accordance with AdS/CFT dictonary this
stress-energy tensor is dual to the metric of spacetime after
collision of shock waves \cite{early-G5}.

The main result of \cite{Gubser,GubserII}, confirmed by numerical
calculations performed in \cite{Shuryak09,DuenasVidal10}, is that
in the limit of a very large collision energy E
the multiplicity (the entropy S) grows as
\be S>{\cal
C}E^{2/3},\ee
${\cal C}$ is a numerical factor (see
Sect.\ref{GYP}).

Alvarez-Gaume, Gomez, Sabeo Vera, Tavanfar, and  Vazquez-Mozoand
\cite{Alvarez} have  considered
 central collision of shock waves sourced by a nontrivial matter
distribution in the transverse space and they  have found critical
phenomenon occurring as the shock wave reaches some diluteness
limit. This criticality  may be related to
criticality found in \cite{Shuryak09}. The numerical results of
\cite{Shuryak09} show the
  existence of a simple scaling relation between the critical impact
  parameter and the energy of colliding waves.

 The size of colliding nuclei is introduced
via the distance of those objects from the boundary along the
holographic coordinate z.

The model of infinite homogenous wall has been proposed and analyzed
by Shuryak and Lin \cite{Shuryak09}. The advantage of this model is
the essential simplicity of calculations. However, the legitimacy of
these calculations requires some additional examinations (see our
discussion in Sect. \ref{Reg-Shyr}).

In heavy ion collisions not only the energy per
nucleus is important
variable. One can try to associate different nuclei with different
kinds of shock waves. There are several proposal
in literature on this subject. For example, in \cite{Gubser08} the holographic
model with cutting off the UV part of the bulk geometry, has been
proposed. Formation of
trapped surfaces (TS) in head-on collisions of charged shock waves in
the (A)dS background has been considered in \cite{ABJ} and it has
been shown
 that the formation of trapped surfaces on the past light cone is only
possible when charge is below certain critical value - situation similar to the collision of two
ultrarelativistic charges in Minkowski space-time \cite{Mann}. This critical value depends on the energy
of colliding particles and the value of a cosmological constant. The formation of trapped surfaces in
head-on collisions of shock waves in gravitational theories with
more complicated bulk dynamics, in particular with the Einstein-dilaton dynamics,
pretended
to describe holographic  physics that is closer
to QCD than the pure  AdS theory \cite{Kiritsis07,Gubser08},
  has been considered recently by Kiritsis and  Taliotis \cite{KirTalio11}\footnote{Collision of
dilatonic shock waves in the flat background has been considered in
\cite{IA-cat}.} and they have found that the multiplicity grows as \be S \gtrsim E^{\,0.24},\ee
that is rather close to the experimental data.

 In this paper we propose to incorporate the study of collisions of charged shock
 gravitational waves \cite{ABJ}
  into the description of colliding nuclei with non-zero baryon chemical potential.
  In the holographic context,
  the chemical potential of strongly coupled QGP on the
   gravity side is related to temporal component $A_t$ of the $U(1)$
   gauge field \cite{chemical}-\cite{Erdmenger}.
The asymptotic value of this gauge field component in
the bulk is interpreted as the chemical potential in the gauge theory
\be
\label{mu-A}
\mu =A_t|_{{\rm boundary}}.\ee
We  use the same identification (\ref{mu-A}) for colliding ions.
It would be interesting to perform
 calculations for the off-center collision of charged gravitational waves
 or generally smeared charged shock waves. Postponing this problem
 for further investigations, here we consider the head-on collision of charged point shock waves
 and charged wall shock waves.
 This will give us the holographic picture for QPG phase diagram in the first moment after collisions of heavy ions.
This  phase diagrams, chemical potential (charge) $\mu$ versus
temperature
 (energy)
$T$, are displayed in Fig. 5 and Fig. 11.
 The colored lines separate the TS-phase from the
phase free of TS. Let us note that the obtained
diagrams differ from the phase diagram for
equilibrium QGP (see Fig.\ref{phaseDiag-07091225} in Sect.
\ref{chen-pot}).

 The paper is organized as follows.
 In Sect.2 we present our set up of the problem.
 In Sect.2.1.1 we describe the role of black holes  in AdS/CFT description of
  strongly coupled QGP. In Sect. 2.1.2. we present the
  description of the chemical potential of QGP within the  AdS/CFT correspondence.
  In Sect. 2.1.3 we remind the main facts about shock waves
  in AdS$_5$ related to the trapped surface formation. In Sect.2.1.4 we describe in
  details the dual conjecture proposed in \cite{Gubser}.
    In Sect.2.2 we pay a special
 attention to the problem of regularization that appears within the wall shock waves approach.
 In Sect. 3 we present  the phase diagram, temperature vs chemical
 potential,
 for QGP formed in the  heavy-ions collisions
 by using the holographic approach with the central collision of charged shock waves.
 In Sect. 4 we present our calculations of the same problem by using the regularized version of the
 charged wall shock waves. We summarize our calculations  in Sect. 5 and present
 here also further directions related to holographic description
  of quark-gluon plasma formed in heavy-ions collisions.
$$\,$$

 \section{Set up}
\subsection{Dual Conjectures}
\subsubsection{Black holes and  AdS/CFT correspondence  for strongly coupled QGP}

The idea of AdS/CFT applications
 to description of the QGP is based on the possibility
to make an one to one correspondence between
phenomenological/termodynamical parameters of plasma -- $T,E,P,\mu$
-- and parameters that characterize deformations of AdS$_5$.  In the
dual gravity setting the source of temperature and entropy are
attributed to the gravitational horizons. The relation between
energy density and temperature typical for the BH in AdS according
\cite{Page,Amanda}  is \be \label{Amanda} E=\frac{3\pi ^3
\,L^3}{16\,G_5}T^4 \ee

In the phenomenological model of QGP, such as the Landau or Bjorken
hydrodynamical models \cite{Landau,bjorken}, the
 plasma is characterized by a
space-time profile of the energy-momentum tensor $
T_{\mu\nu}(x^\rho),\,\mu, \nu, \rho =0,...3. $ This state has its
counterpart on the gravity side as a modification of the geometry
of the original AdS$_5$ metric. This follows  the general AdS/CFT
line: operators in the gauge theory correspond to
fields in SUGRA.
 In the case of the energy-momentum tensor,
the corresponding field is just the 5D metric. It is convenient to
parameterize corresponding 5-dimensional geometry as
\be
ds^2=L^2\frac{g_{\mu\nu}(x^\rho,z) dx^\mu dx^\nu +dz^2}{z^2},
\label{Fefferman-Graham}
\ee
that is the 5D Fefferman-Graham
metric \cite{fg}. The flat case $g_{\mu\nu} = \eta_{\mu\nu}$
parametrizes AdS$_5$ in Poincar\'e coordinates. The conformal
boundary of space-time is at $z\!=\!0$ and \be\label{bci}
g_{\mu\nu}(x^\rho,z)=\eta_{\mu\nu}+z^4
g^{(4)}_{\mu\nu}(x^\rho)+\ldots \ee
The AdS/CFT duality
leads to the relation \be \label{T-g}
g^{(4)}_{\mu\nu}(x^\rho) \sim <T_{\mu\nu}(x^\rho)> \ee

Applications of AdS/CFT correspondence to
hydrodynamical description of the GQP
 is based on the fact that
  the energy momentum tensor can be read off from the
expansion of the BH in $AdS_5$ metric (\ref{bci})
 corresponding to simple hydrodynamical model
\be <{T_{\mu\nu}}>\propto g^{(4)}_{\mu\nu} ={\mbox
{diag}}(3/z_0^4,1/z_0^4,1/z_0^4,1/z_0^4) \ee The  BH in $AdS_5$ in
the  Fefferman-Graham coordinates has the form
(\ref{Fefferman-Graham}) with the following nonzero components of
$g_{\mu\nu}(x^\rho,z)$ (see \cite{HoQGP-review} and refs therein)
\bea \label{e.bhfef}
g_{00}=-\frac{\left(1\!-\!\frac{z^4}{z_0^4}\right)^2}{\left(1\!+\!\frac{z^4}{z_0^4}\right)},\,\,\,\,\,\,\,
g_{ii}=  \left(1\!+\!\frac{z^4}{z_0^4}\right) \eea A change of
coordinates $ \tilde z = z/(1+z^4/z_0^4)^{1/2} $ transforms
(\ref{Fefferman-Graham}) to the standard metric form of the
AdS-Schwarzschild static black hole
\bea
\tilde
z^2 ds^2=-\left(1-\frac{\tilde z^4}{\tilde z_0^4}\right) dt^2\! +
d\vec x^2\! + \frac{1}{1-\frac{\tilde z^4}{\tilde z_0^4}} d\tilde
z^2, \label{standard}
\eea
with $\tilde z_0=z_0/\sqrt{2}$ being the
location of the horizon.

\subsubsection{Chemical potential in  QGP via AdS/CFT correspondence}
\label{chen-pot}

The Reissner-Nordstr\"om metric in
AdS has
 the following form:
\begin{equation}
\label{RN-lambda} ds^2=-g(R)dT^2+g(R)^{-1}dR^2+R^2d\Omega_{D-2}^2,
\end{equation}
\begin{equation}
\label{g}
g(R)=1-\frac{2M}{R^{2}}+\frac{Q^2}{R^{4}}+\frac{\Lambda}{3}R^2,\end{equation}
where $\Lambda$  is cosmological constant, $\Lambda/3\equiv
1/a^2$, $M$ and $Q$ are related to the ADM (Arnowitt-Deser-Misner)
mass $m$ and the charge $\sigma $
\begin{equation} M=\frac{4\pi G_5m}{3\pi^2},\quad Q^2=\frac{4\pi G_5\sigma^2}{3}.
\end{equation}
$\sigma$ is a charge of the electromagnetic field (pure electric)
with only one  non-zero component \be \label{At}
A=A_TdT=\Big(-\sqrt{\frac{3}{4}}\frac{ Q}{R^{2}}+\Phi \Big) dT,
\ee here $\Phi$ is a constant $
\Phi=\displaystyle\frac{\sqrt{3}}{2}\displaystyle\frac{
Q}{R_+^{2}},$ where $R_+$ is the largest real root of $g(R)$.
Thermodynamics of the charged BH is described by the grand
canonical potential (free energy) $W=I/\beta$, the Hawking
temperature $T=1/\beta$,
 and the chemical
potential \cite{charge,charge1} that are given by \be
I=\frac{\pi\beta}{8G_5a^2}\left(a^2R_+^2+R_+^4-\frac{Q^2a^2}{R_+^2}\right),\,\,\,
T=\frac{1}{4\pi}g'(R_+),\,\,\,\,\mu =\frac{\sqrt{3}Q}{2R_+^2},\ee
here $R_+$ is outer horizon, $g(R_+)=0$, $I$ is given by the value
of the action at (\ref{g}) and (\ref{At}). The relation with the
first low of thermodynamics, $d{\cal E}=Td{\cal S}+\mu d{\cal Q}$
is achieved under identification \bea W={\cal E}-T{\cal S}-\Phi
{\cal Q},\,\,\,\, {\cal E}=m,\,\,\,{\cal
S}=\frac{S_H}{4G_5},\,\,\,\,{\cal Q}=q,\,\,\,\mu=\Phi \eea Note
that just the asymptotic value of a single gauge field component
gives the chemical potential \cite{chemical}-\cite{Erdmenger} \be
\mu=\lim_{r\to \infty}A_t(r)\label{mu}\ee

The QGP is characterized at least by two parameters: temperature and chemical
  potential. Generically speaking, quantum field theories may  have non zero chemical potentials for any or
all of their Noether charges. Within the AdS/CFT context
two different types of chemical potential are  considered, namely related to the R charge and to baryon number.

Baryon number charge can only occur when we have a theory containing fundamental
flavours.  Introduction
flavours into the gauge theory by means of a D7 brane leads to appearance of a $U(N_f )$ global
flavour symmetry. The
flavour group contains a $U(1)_B$, that is a baryon number symmetry, and for this baryon number
one adds  a chemical potential $\mu _b$ \cite{findens}. To calculate the free energy
 one has to calculate the  DBI action for a D7 brane. Note that there  is a divergence in formal definition,
 so we must go through the process of renormalization, see for example lectures \cite{Sk} and for yearly discussion
\cite{AV-ren}.

 R charge chemical potential appears for SUSY models \cite{Erdmenger}.
In the N = 1 case there is a U(1) R symmetry group. As to extended SUSY, say N=2, the
quark mass term breaks R symmetry.

 \begin{figure}[h] \centering
 \includegraphics[height=4cm]{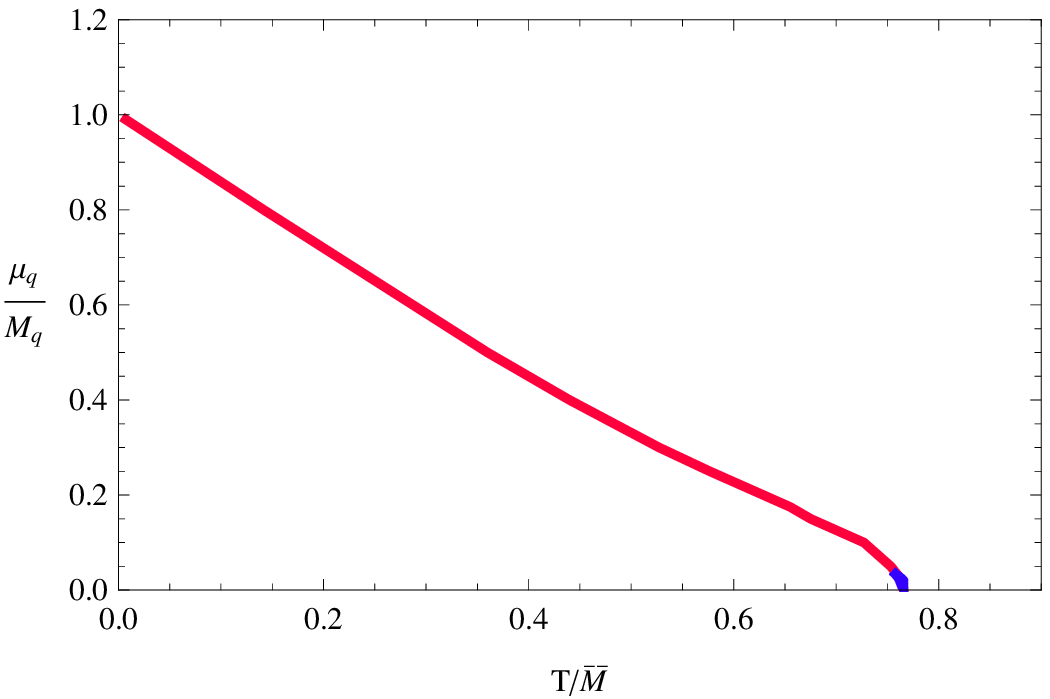} A.$\,\,\,\,\,$
 \includegraphics[height=4cm]{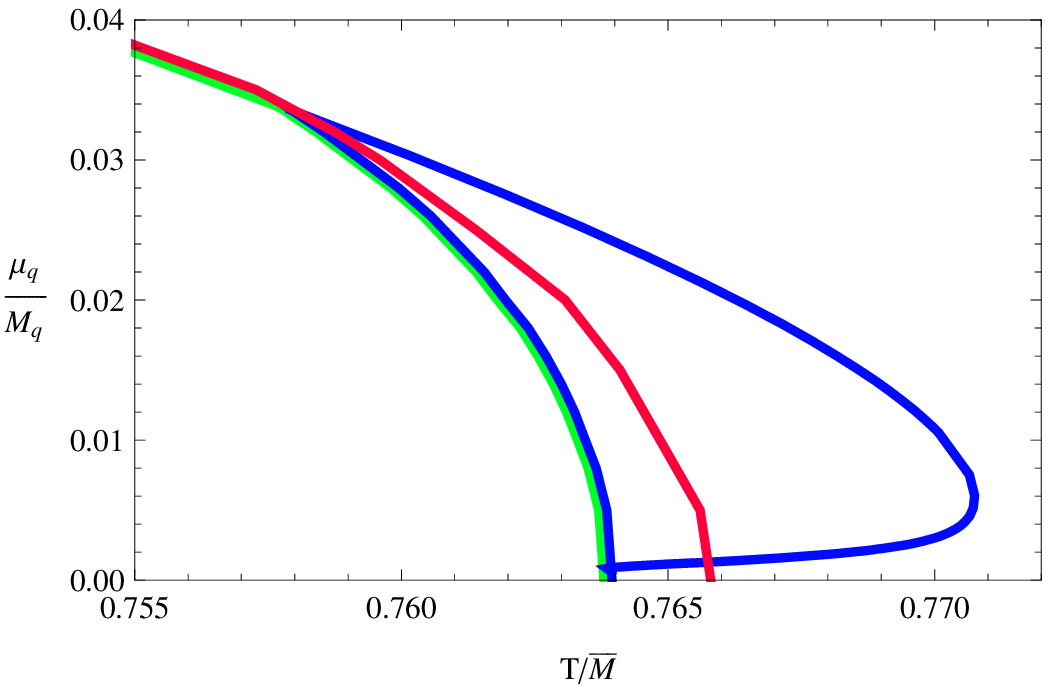}B.
\caption{Phase diagram from \cite{chemical}: Quark chemical potential $\mu_q/M_q$,
in versus
temperature $T/{\bar M}$. The red line separates the phase of Minkowski
embeddings (small temperatures, small $\mu_q/M_q$) from black hole
embeddings (see details in \cite{chemical}). Figure (b) zooms in on the region near the end of this
line. Different lines in B. correspond to different embedding geometries.
}\label{phaseDiag-07091225}
\end{figure}

The typical phase diagram the chemical potential vs the
temperature is presented in Fig. \ref{phaseDiag-07091225} (the
diagram is taken from \cite{chemical}). In the phase diagram:
$\mu_q = \displaystyle\frac{\mu_b}{N_c}$, $\mu_q$ is the quark
chemical potential and $\bar M \propto m_q$ is a mass scale
defined as $\bar M=2M_q/\sqrt{\lambda}$ and $\lambda = g_{YM}^2
N_c $.

 \subsubsection{ Shock waves in AdS$_5$}
Shock
waves propagating in AdS have the form
\be
ds^2=L^2\frac{-dudv+dx_{\perp}^2+\phi(x_{\perp},z)\delta(u)du^2+dz^2}{z^2},
\label{shock}\ee
where $u$ and $v$ are light-cone coordinates, and
$x_{\perp}$ are coordinates transversal to the direction of motion
of the shock wave and to $z$-direction.
This metric is sourced by the stress-energy momentum tensor $T_{MN}$
with only one non-zero component
$T^{SW}_{uu}$
\be
\label{T-J}T^{SW}_{uu}=J_{uu}(z,x_\bot)\delta(u)
         \ee
and the Einstein E.O.M.
reduces to
\be
\label{J-z-bot}
(\Box_{H_3}-\frac3{L^2})\Phi(z,x_\bot)=-16\pi G_5 \frac{z}{L}J_{uu}(z,x_\bot)\ee
where
\be
\label{Phi-phi}
\Phi(z,x_\bot)\equiv\frac{L}{z}\phi(z,x_\bot)\ee
and
\be
\label{Box-z}
\Box_{H_3}=\frac{z^3}{L^2}\frac{\partial}{\partial z }z^{-1}\frac{\partial}{\partial z }
+\frac{z^2}{L^2}(\frac{\partial^2}{\partial x_\bot ^2 })\ee

  Different forms of the shock waves correspond to different forms of the source
$J_{uu}(z,x_\bot)$.
The most general $O(3)$ invariant shock  wave in AdS located at $u=0$
        corresponds to \be
\label{O3}
\Phi^{O(3)}(z,x_\bot)=F(q).
\ee
where   $q$ is the chordal distance
 \be
q=\frac{(x^1_{\perp})^2+(x^2_{\perp})^2+(z-z_0)^2}{4zz_0},\ee
 In  this case  $\rho$, related to $J_{uu}$ as
        \be
\label{rho}
 \frac{z}{L}J_{uu}(z,x_\bot)\equiv \rho (z,x_\bot),
\ee has the form
        \be
        \rho^{O(3)}(z,x_\bot)=\rho(q),\ee
 and the Einstein E.O.M takes the form
\be
   (\Box _{H_3}-\frac{3}{L^2})F=-16\pi G_5 \,\rho(q)
   \ee
   or explicitly\be
\label{stress-energyequation}
q(q+1)F''_{qq}+(3/2)(1+2q)F'_q-3F=-16\pi G_5 L^2\,\rho(q),
   \ee

 The shape of point shock wave   $F^{p}$
 is given by the solution to (\ref{J-z-bot}) with
\be\label{point}
J_{uu}=E\delta(u)\delta(z-L)\delta(x^1)\delta(x^2)
\ee
and has the form is given by
\be
\label{Phi-point}
F^{p}(z,x_\bot)=\frac{8L^2G_5Ez^3}{(x_\bot^2+(z-L)^2)^3}
\ee
This  point shock wave shape is in fact equal to $F^p(q)$,
$\Phi^{point}(z,x_\bot)=F^p(q)$, that is a solution to
(\ref{stress-energyequation}) with
 \be
\label{psw-rho}
\rho^p(q)=
\frac{E}{L^3} \,\frac{\delta(q)}{\sqrt{q(1+q)}}.\ee
It has the form
\be
 \label{Fp}
  F^p=\frac{2G_5E}{L}\,\left(\frac{(8q^2+8q+1)-4(2q+1)\sqrt{q(1+q)}}{\sqrt{q(1+q)}}\right)\\
\ee

The shape of the charged point shock wave is a sum of two components
\be
 \label{F}
  F=F^p+F^Q
  \ee where $F^p$ is given by (\ref{Fp}) and $F^Q$ is the solution to
(\ref{stress-energyequation}) with
 \be
 \label{rho-charge}
 \rho^{pQ}=\frac{5\bar{Q}^2}{32\cdot 64\, L^5G_5}\frac{1}{[q(q+1)]^{5/2}}=
\frac{5Q_n^2}{\pi 24\cdot 64\, L^5}\,\frac{1}{[q(q+1)]^{5/2}} ,\ee
explicitly
\be
\label{FQ}
F^Q=\frac {5G_5 Q_n^2
}{48{L}^{3}}\,\frac{144q^2+16q-1+128q^4+256q^3-64(2q+1)q(q+1)\sqrt{q(1+q)}}{q(1+q)\sqrt{q(1+q)}}
\ee

To establish the connection with \cite{ABJ} let us note the relations of notations
\be
\label{Mbar}\bar{M}=\frac{4G_5E}{3\pi}\ee
\be
\label{Qbar}\bar Q^2=
\frac{4G_5 Q_n^2}{3\pi}
\ee and
\bea
\frac{3\pi \bar M}{2a}|_{\,{\mbox {notations of}}\,\cite{ABJ}}&=&\frac{2G_5E}{L}
|_{\,{\mbox {notations of}}\,\cite{Gubser}\, \mbox{and here}}\\
\frac{5\pi \bar Q ^2}{64 a^3}|_{{\,\mbox {notations of}}\,\cite{ABJ}}&=&\frac {5G_5 Q_n^2
}{48{L}^{3}}
|_{{\mbox {notations here}}}\eea

More complicated shock waves in AdS and dS have been considered  in
\cite{Hotta,Sfetsos,Horowitz,Emparan,SHOCK-ADS-GEN,ABG}.

\subsubsection{GYP Dual Conjecture}\label{GYP}

Gubser, Yarom and
 Pufu  (GYP) \cite{Gubser} have proposed the following dual to QCD holographic picture for
colliding nuclei:
\begin{itemize}
\item
the bulk dual of the boundary nuclei is the shock
waves propagating in AdS of the form (\ref{shock});
\item
the bulk dual of two colliding
nuclei in the bulk is the line element for two identical
 shock waves propagating towards one another in AdS
\be
ds^2=L^2\frac{-dudv+dx_{\perp}^2+\phi_{1}(x_{\perp},z)\delta(u)du^2+
+\phi_{2}(x_{\perp},z)\delta(v)dv^2+dz^2}{z^2};
\label{two-shock}\ee
\item when the shock waves collide in the bulk, a black hole should form,
 signifying the formation of a quark-gluon-plasma.
\end{itemize}

 To estimate the BH formation one usually use the TS technic \cite{Penrose,EG}
 \footnote{This estimation can be also performed using so-called capture arguments
 \cite{Kaloper,IA-cat}.}.
 A trapped surface is a surface whose null
normals all propagate inward \cite{HawkingEllis}.
 There is no rigorous proof that the  TS formation in asymptotically AdS spacetime provides the BH formation,
 however there is an
common belief  that trapped surfaces must lie behind an event horizon and that
a lower
bound on entropy $S_{AdS}$ of the black hole is given by
 area of the TS, $A_{ trapped}$,
\be
 \label{PB}
  S_{AdS} \geq S_{ trapped} \equiv \frac{A_{ trapped} }{ 4 G_5}
\ee

To make the proposed duality prescription more precise one has to fix the
relations between the bulk parameters, $G_5,L,E$
and the phenomenological parameters of QGP.
According to \cite{Gubser08}, one of these relations is
 \be
 \label{LG5}
 \frac{ \,L^3}{G_5}=\frac{16E\cdot T^4}{3\pi ^3}=\frac{11\cdot16}{3\pi ^3}\approx 1.9\ee
The arguments supporting  (\ref{LG5}) are following.
Lattice calculations in QGP  \cite{Lattice} have shown that $ET^4$
 is a slowly varied quantity and
\be \label{lattice-cal}
{E}{T^4}\approx 11.\ee
Just to match the black hole equation of state (\ref{Amanda})
with (\ref{lattice-cal}), Gubser, Yarom and
 Pufu \cite{Gubser} have assumed (\ref{LG5}).
 It is important to note that here is assumed
an identification of the total energy of each nucleus with the energy of the
   corresponding shock wave. One can modify this identification and assume
   that only a part of energy  of the gravitational shock wave is related
   to the total energy of nucleus.

   To fix the dimensioòless parameter $EL$ the  AdS/CFT  dual relation
   (\ref{T-g})
   between
the expectation value of the gauge theory stress tensor
and the $AdS_5$ metric deformation by  the  shock wave has been used
\cite{Gubser},

\be \label{CTuu}
  \langle T_{uu}  \rangle =
   {L^2 \over 4\pi G_5} \lim_{z\to 0}
   {1\over z^3} \Phi(z,x_\bot) \delta(u)
 \ee
For the point shock wave
 $\Phi^{point}$  given by (\ref{Phi-point}),
one gets the following
stress tensor in the boundary field theory:
\be
\label{CPTuu}
\langle T_{uu}  \rangle =\frac{2L^4E}{\pi(L^2+(x^1)^2+(x^2)^2)^3}\delta(u)
\ee
The  RHS of (\ref{CPTuu})  depends on  the total energy $E$ and $L$, and $L$
has a meaning  of the root-mean-square radius of the
transverse energy distribution.
It has been assumed \cite{Gubser} that
  $L$ is equal to the root-mean-square transverse radius of the nucleons,
  that is in according with
   a Woods-Saxon profile for the nuclear density
   \cite{Klein:1999qj,Adams:2004rz} is of order of few fm. In particular for Au
  it is equal to
   $L \approx 4.3\,{\rm fm}$. For Pb it is $L \approx 4.4\,{\rm fm}$.

The RHIC collides Au nuclei, (A=197), at $\sqrt{s_{NN}} = 200\,{\rm GeV}$.
This means that each nucleus has energy $E = 100\,{\rm GeV}$
 per nucleon, for a total of about $E = E_{\rm beam} =19.7\,{\rm TeV}$ for each nucleus.

 LHC will collide  Pb nuclei, (A=208)
at $\sqrt{s_{NN}} = 5.5\,{\rm TeV}$, that means
$E = E_{\rm beam} = 570\,{\rm TeV}$.

Estimations of \cite{Gubser} for
dimensionless values $ EL $ for Au-Au and Pb-Pb collisions are
 \bea\label{FoundEL}
   EL |_{Au-Au,\sqrt{s_{NN}} = 200\,{\rm GeV}}&\approx& 4.3 \times 10^5 \,,
\\
   EL |_{Au-Au,\sqrt{s_{NN}} = 5.5\,{\rm TeV}}&\approx& 1.27 \times 10^7 \,,
   \eea

Note, that in \cite{Shuryak09} has been proposed to  tune  the
scale $L$ or $z_0$ of the bulk colliding object to the size of the
nucleus, or to the ``saturation
scale'' $Q_s$ in the
 ``color glass'' models.

Calculations in \cite{Gubser} show that  in the limit of a very large collision
energy E  the entropy grows as $E^{2/3}$,
\be\label{STE}
  S_{\rm trapped} \approx \pi \left( {L^3 \over G_5} \right)^{1/3}
    (2EL)^{2/3},
    \ee
Considerations of off-center collisions
of gravitational shock waves in AdS do not change the scaling $E^{2/3}$.
However, a critical impact parameter, beyond
which the trapped surface does not exist has been observed \cite{Shuryak09}
(compare with result of
\cite{Alvarez}). Experimental indications for similar
critical impact parameter in real collisions have been noted \cite{Shuryak09}.

The relation of the total multiplicity, $S_{QGP}$, given by experimental data, and the
entropy produced in the gravitational waves collision in $AdS_5$,
$S_{AdS}$ has some subtleties
\cite{Gubser08}.
Phenomenological considerations  \cite{Ochs,Muller,Gubser},
estimate  the total multiplicity $S_{QGP}$ by the  the number of
charged particles $N_{ch}$ times the factor $\sim 7.5$.
\be
\label{M-N}
  S_{QGP} \approx 7.5 N_{\rm charged} .
 \ee

The trapped surface analysis does not give the produced entropy but it
provides a lower bound
\be \label{S-bound}
S_{trapped}\leq S_{AdS}.
\ee

\begin{figure}[ht] \centering
 \includegraphics[height=5cm]{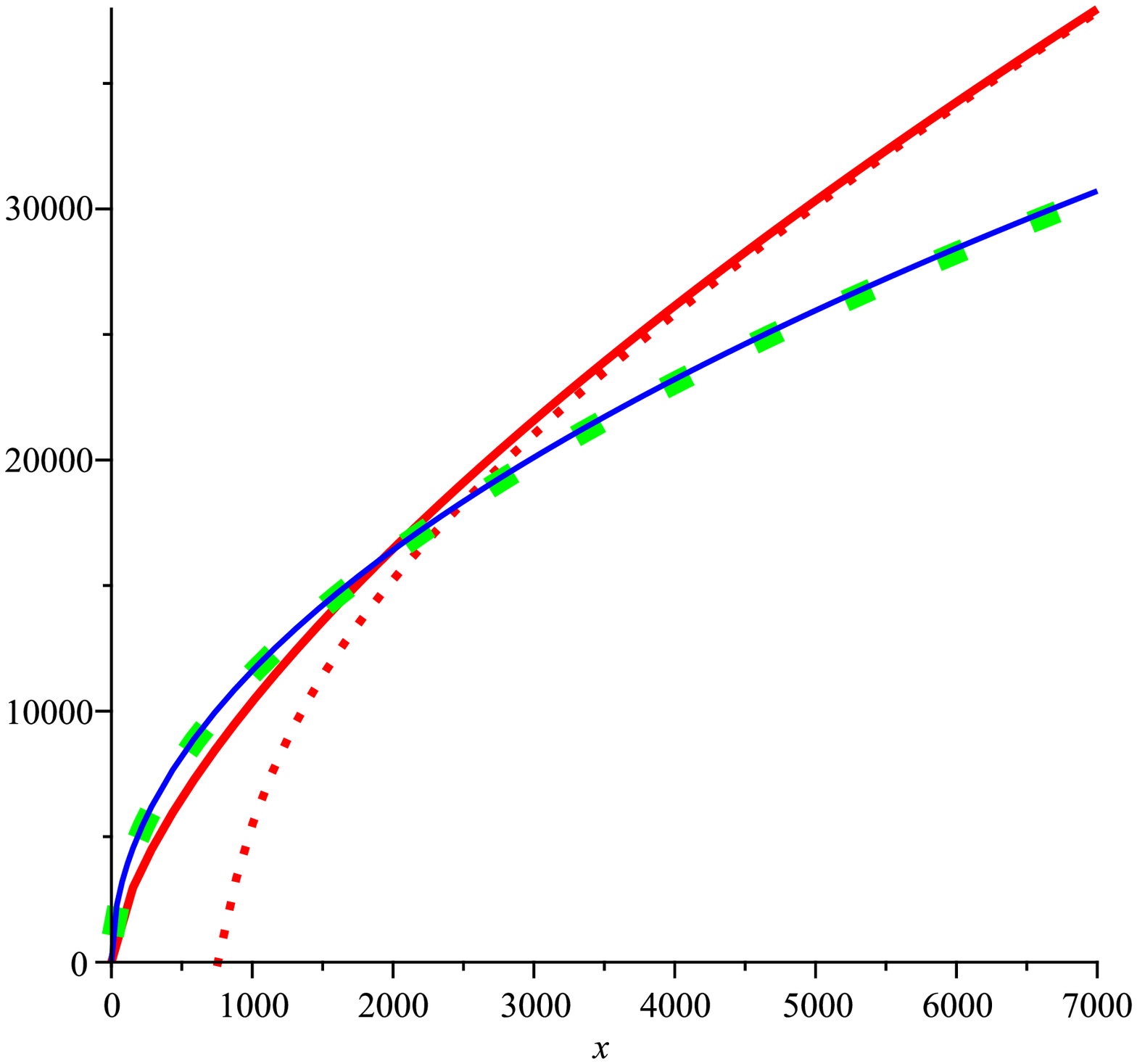}$\,\,\,\,\,$
 \includegraphics[height=5cm]{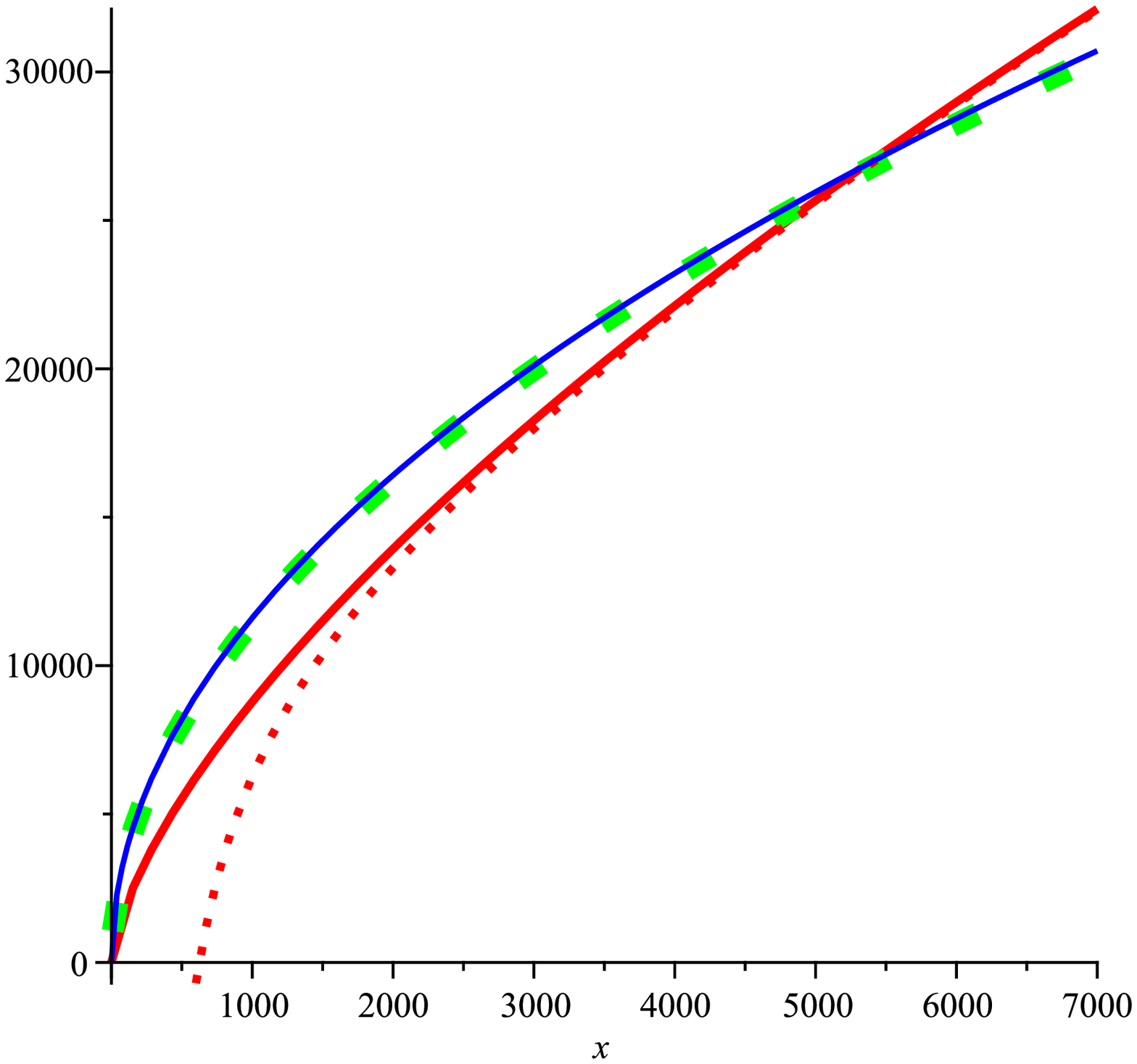}
  \caption{
  (color on-line) Plots of the total number of charged particles
vs.~energy. The red lines present the estimation (2.45).
Plots A and B differ by the
 overall factor ${\cal M}$.
The blue lines correspond to the prediction of the Landau model
and the dotted green lines schematically present the curves that fit experimental data.
The dashed lines correspond to corrections to the GYP multiplicity
via non-zero chemical potential, see Sect.3.}
\label{CorGYP}
 \end{figure}

Taking into account that in calculations \cite{Gubser} the energy of the gravitational shock wave
is identified with the
energy of colliding ions and $L$ with the
nucleus size, one can introduce proportionality constants between
these quantities to get
\be
{\cal M}\cdot S_{trapped}<N_{\rm charged}\ee
where all  proportionality  factors are included into the  overall factor ${\cal M}$.
One can take  ${\cal M}$  to fit the experimental data at some point.
But  the scaling $S_{\rm trapped} \propto
s_{NN}^{1/3}$ implied by (\ref{STE}) differs from the observed
scaling, which is closer to the dependence $S \propto s_{NN}^{1/4}$, that
predicted by the Landau model \cite{Landau}, see Fig.\ref{CorGYP}.
It is obvious, that if  $E<E_{max}$ one can avoid
a conflict between \cite{Gubser} and
experiment, but if E can be arbitrary large
the conflict takes place.

In figure \ref{CorGYP} we  plot the dependence of the
entropy bound (\ref{STE}) on the energy, together with
the curve that schematically represents the realistic curve that fits experimental
data  \cite{Back:2002wb}. We can see that by changing the coefficient ${\cal M}$
one can avoid the conflict only for energy up to some $E_{max}.$
The overall coefficient of
the numerical plot has been chosen in order to fit the RHIC data \cite{Back:2002wb}.
Their are
indicated by dots
in Fig.\ref{CorGYP}.

In the above estimation
 energy of each shock wave is identified with  the energy of colliding beams.
As has been noted in \cite{GubserII} one can  improve fit to the data by identifying the
 energy of each shock wave with the fraction of the energy of the nucleus
 carried by nucleus that participate in the collision.
 This give an extra parameter to fit data.
But still a conflict will arise at large energies. In paper \cite{GubserII}
it has been proposed
to cure the problem by removing a UV part of AdS bulk. In \cite{KirTalio11}
shock waves corresponding to the BH with non-zero dilaton field \cite{Kiritsis07}
were considered and it has been shown that
lower bound on $N_{\rm charged}$ scales is rather closer to  $s_{NN}^{1/4}$.


\subsection{Remarks about the regularization of TS calculations in the case of
wall-on-wall collisions } \label{Reg-Shyr}

In \cite{Shuryak09} has been proposed  a much simpler dual description
of the colliding
nuclei that uses a wall-on-wall collision
in the bulk.
The Einstein equation  for the profile of the wall shock wave \cite{Shuryak09} has the
form:
\begin{eqnarray}\label{EinsteinwithoutQ}
         (\partial^2_z-\frac{3}{z}\partial_z)\phi(z)=J^{WP}_{uu},\,\,\,\,J^{WP}_{uu}=-16\pi
         G_5\displaystyle\frac{E}{L^2}\frac{z_0^3}{L^3}\delta(z-z_0)
         \end{eqnarray}
To find a trapped surface that can be formed in the collision of two wall shock waves
 one needs to find a solution to the
Einstein eq.(\ref{EinsteinwithoutQ}) that satisfies two conditions.
It is convenient to write these conditions in terms of function $\psi(z)$ related to $\phi$
via
\be\phi(z)=\displaystyle\frac{z}{L}\psi.\ee
They have the form
\begin{eqnarray}
  &&\psi(z_a)=\psi(z_b)=0, \label{boundary-f}\\
  &&\psi^{\prime}(z_a)\displaystyle\frac{z_a}{L}=2,\qquad
  \psi^{\prime}(z_b)\displaystyle\frac{z_b}{L}=-2\label{boundary-d}
\end{eqnarray}
where $z_a,$ $z_b$ are supposed to be the boundaries of the trapped
surface \cite{Shuryak09}.  But as we will see in the moment, strictly speaking,
one cannot call the solution to the equation (\ref{EinsteinwithoutQ}) with b.c.
(\ref{boundary-f})
and (\ref{boundary-d}) the trapped surface, since by definition this surface supposed to be
smooth and compact meanwhile  the solution \cite{Shuryak09} is {\it non-smooth} and {\it noncompact}.

By this reason we call  the solution found in \cite{Shuryak09} a quasi-trapped
surface. Let us remind the construction presented in \cite{Shuryak09}.

 In \cite{Shuryak11}, the
solution to the Einstein equation (\ref{EinsteinwithoutQ}) is
written in such a way that the property \eqref{boundary-f} is
satisfied automatically. This solution has the form
\be\psi(z)=\psi_a(z)\Theta(z_0-z)+\psi_b(z)\Theta(z-z_0)
\label{solution 11-i}\ee
$$\psi_a(z)=-\frac{4\,G\pi \,E\left( {\displaystyle\frac {z_0^{4}}{{{z_b}}^{4}}}-1
 \right) {{z_b}}^{4}{{z_a}}^{3} \left( {\displaystyle\frac {{z}^{3}}{{{z_a
}}^{3}}}-{\frac {{z_a}}{z}} \right)}{{L}^{4} \left( {{z_b}}^{4}-
{{z_a}}^{4} \right)}$$
$$\psi_b(z)=-\frac{4\,G\pi \,E \left( {\displaystyle\frac {{{z_0}}^{4}}{{{z_a}}^{4}}}-1
 \right) {{z_a}}^{4}{{z_b}}^{3} \left( {\displaystyle\frac {{z}^{3}}{{{z_b
}}^{3}}}-{\frac {{z_b}}{z}} \right)}{{L}^{4} \left( {{z_b}}^{4}-
{{z_a}}^{4} \right)}$$ \\
\noindent

Let us first note that solution \eqref{solution 11-i} is not
smooth.
There is a non-smooth part of the solution \eqref{solution 11-i}
\begin{eqnarray}
  &&\Xi=\frac{{\cal K}}{z}\left(-\frac{z_b}{z_a^3}\left(\Upsilon_1\right)-\frac{z_a}{z_b^3}\left(\Upsilon_2\right)\right), \quad\mbox{ãäå} \\
  &&\Upsilon_1= z^4\Theta(z_0-z)+z_0^4\Theta(z-z_0)\label{Upsilon1m}\\
  &&\Upsilon_2=z_0^4\Theta(z_0-z)+z^4\Theta(z-z_0)\label{Upsilon2m}
\end{eqnarray}
where
\be {\cal K}=\frac{4\,G\pi \,E}{L^4}\frac{z_a^3z_b^3}{z_b^4-z_a^4}
\ee Thus, in order to smooth the solution we have to
smooth the function $\Xi.$ We can do it by performing the regularization
of the Heaviside step function
\begin{eqnarray}
  \Theta(z_0-z)&\approx&\Gamma_1={\frac {\arctan \left( R
\left( z_0-z\right) \right)^3 }{\pi }}+ \frac{1}{2} \label{theta(z0-z)}\\
  \Theta(z-z_0)&\approx& \Gamma_2={\frac {\arctan \left( R \left(z-z_0\right)  \right)^3 }{\pi
}}+\frac{1}{2} \label{theta(z-z0)}:
\end{eqnarray}
and considering the regularized functions
$\tilde{\Upsilon}_1$ and $\tilde{\Upsilon}_2$
\begin{eqnarray}
  &&\tilde{\Upsilon}_1= z^4\left({\frac {\arctan \left( R
\left( z_0-z\right) \right)^3 }{\pi }}+
\frac{1}{2}\right)+z_0^4\left({\frac {\arctan \left( R
\left(z-z_0\right)  \right)^3 }{\pi
}}+\frac{1}{2}\right) \label{Upsilon1}\\
  &&\tilde{\Upsilon}_2=z_0^4\left({\frac {\arctan \left( R
\left( z_0-z\right) \right)^3 }{\pi }}+
\frac{1}{2}\right)+z^4\left({\frac {\arctan \left( R
\left(z-z_0\right) \right)^3 }{\pi }}+\frac{1}{2}\right)
\label{Upsilon2}
\end{eqnarray}

For derivatives we have
\begin{eqnarray}&&\displaystyle\frac{d{\Upsilon}_1}{dz}\approx4{z}^{3}\Theta(z_0-z),\quad\displaystyle\frac{d\tilde{\Upsilon}_1}{dz}\approx\displaystyle{
\frac{4{z}^{3}\left( \arctan (R(z_0-z))^{3}+\pi\right)}{\pi}};
\label{der-Up-1}\\
&&\displaystyle\frac{d{\Upsilon}_2}{dz}\approx4{z}^{3}\Theta(z-z_0),\quad
\displaystyle\frac{d\tilde{\Upsilon}_2}{dz}\approx \,{\frac
{4{z}^{3} \left(\arctan (R(z-z_0))^{3} +\pi\right) }{\pi }}.
\label{der-Up-2}\end{eqnarray}

\begin{figure}[ht] \centering
 \includegraphics[height=4cm]{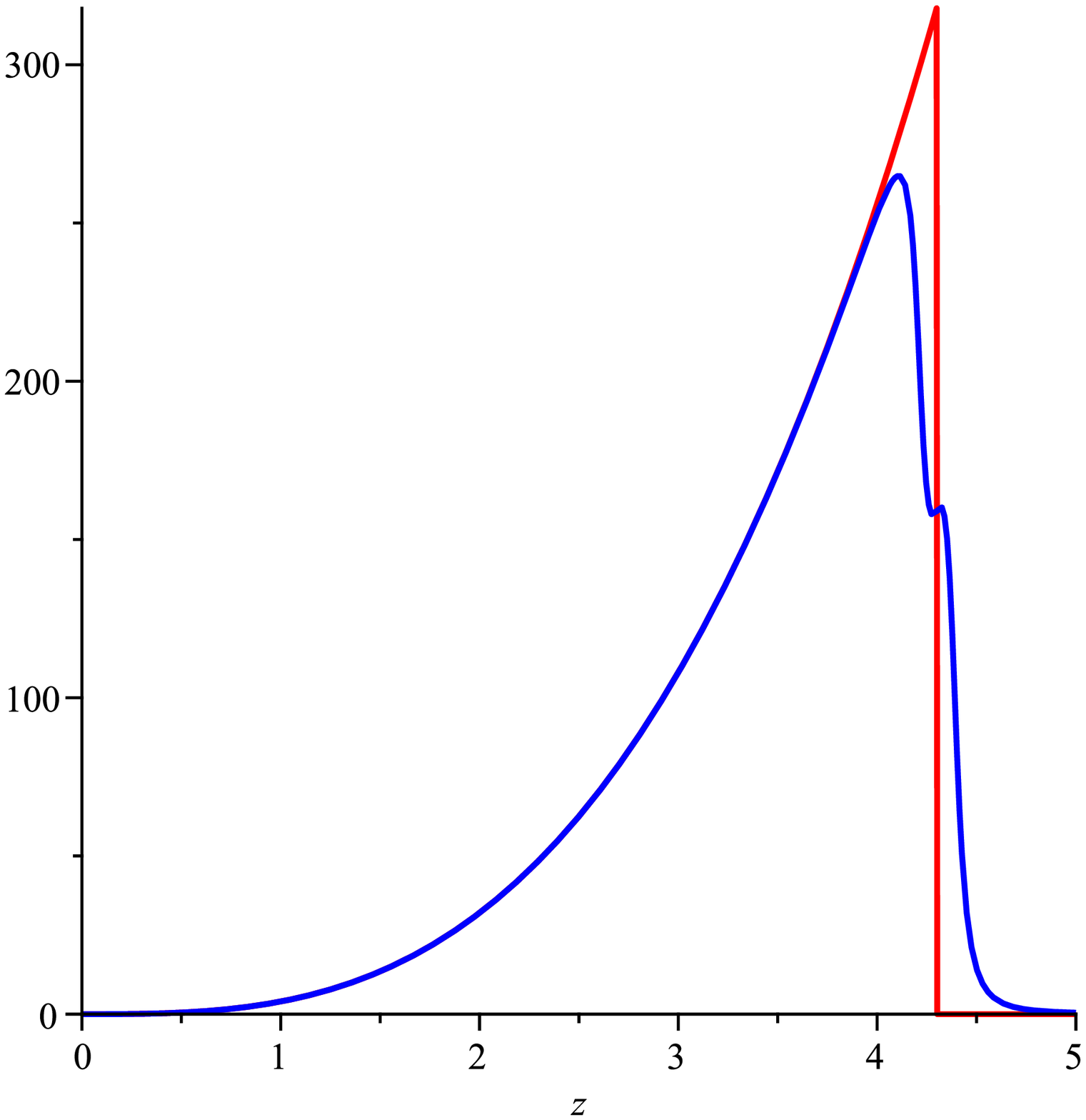}$A.\,\,\,\,\,\,\,$
 \includegraphics[height=4cm]{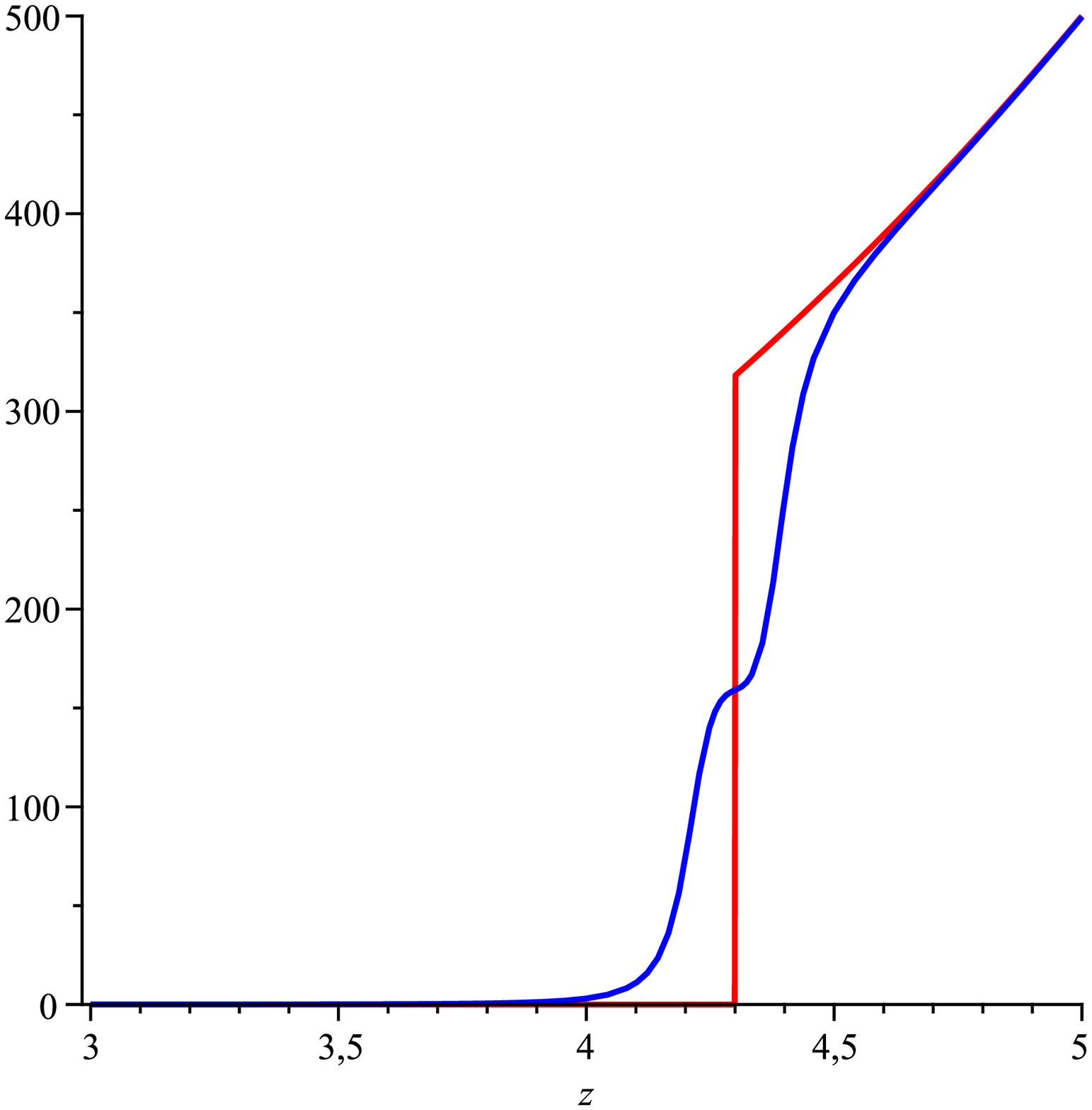}$B\,\,\,\,\,\,\,\,\,$
 \includegraphics[height=4cm]{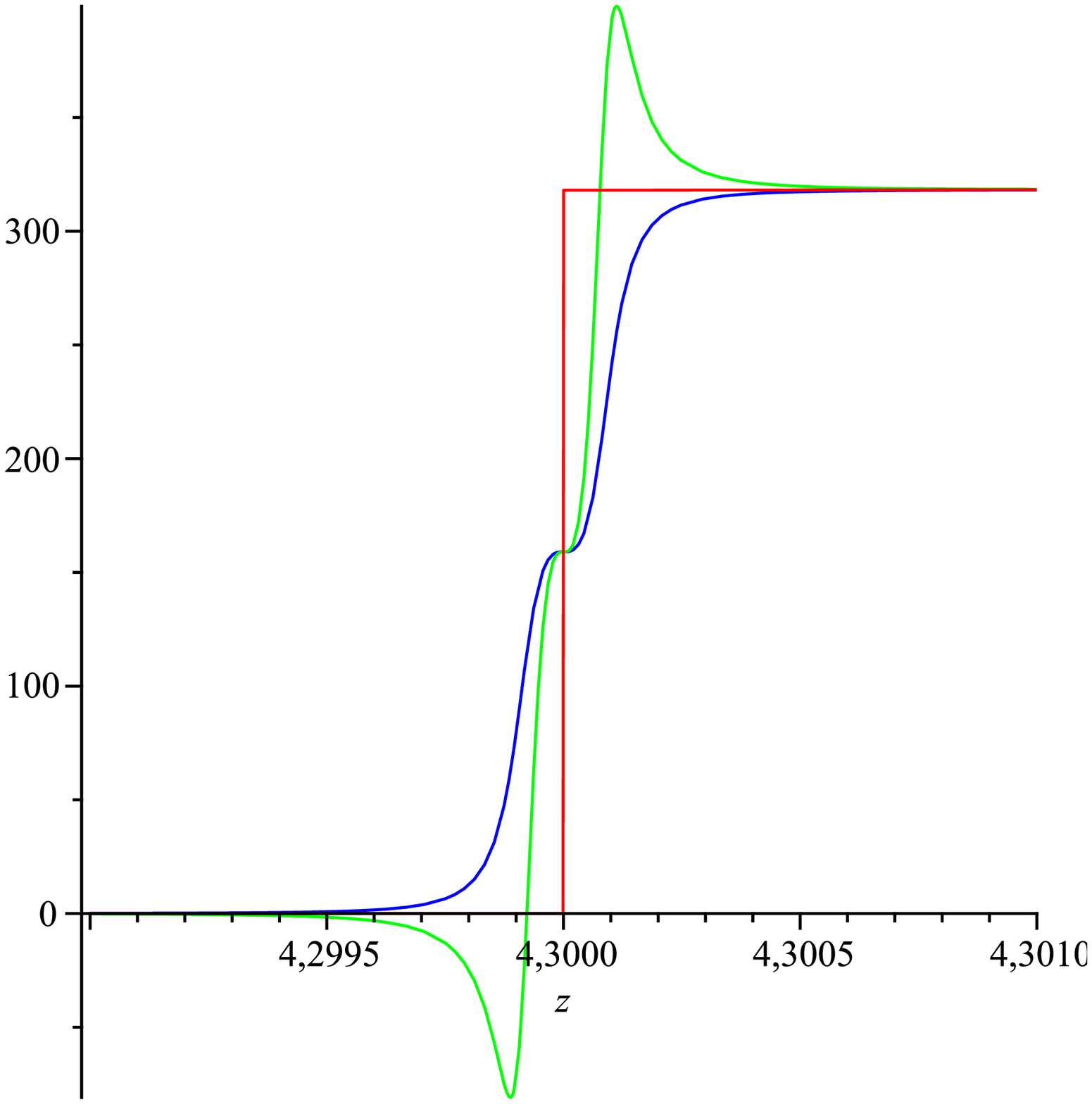}C.
\caption{ A. The functions
$\displaystyle\frac{d{\Upsilon}_1}{dz}$(red line),
$\left.\displaystyle\frac{d\widetilde{
\Upsilon}_1}{dz}\right|_{appr}$ (blue line) . B. The functions
$\displaystyle\frac{d{\Upsilon}_2}{dz}$ (red line),
$\left.\displaystyle\frac{d\widetilde{
\Upsilon}_2}{dz}\right|_{appr}$ (blue line). The regularization
parameter $R=10$ at A and B cases. $\,\,\,\,\,\,\,\,\,$C. Functions
$\displaystyle\frac{d{\Upsilon}_2}{dz}$(red line),
$\left.\displaystyle\frac{d\widetilde{
\Upsilon}_2}{dz}\right|_{appr}$ (blue line) and
$\displaystyle\frac{d\widetilde{ \Upsilon}_2}{dz}$ (green line) at the
regularization parameter $R=10^4$.}
\label{dYBI}\end{figure}

In Fig.\ref{dYBI} we present   the derivatives of functions
$\Upsilon_1,$ $\Upsilon_2$
as well as derivatives of the
smoothed  functions $\tilde{\Upsilon}_1,$ $\tilde{\Upsilon}_2$.

 For
$R=10^4$ (see below) the  differences between derivatives
$\displaystyle\frac{d\widetilde{\Upsilon}_i}{dz}$ and their
approximations given by (\ref{der-Up-1}) and (\ref{der-Up-1})
 \bea\Delta _1(z)&=&\frac{d\widetilde{\Upsilon}_1}{dz}-\left(\frac{d\widetilde{
\Upsilon}_1}{dz}\right)_{appr},\,\,\,\, \Delta
_2(z)=\frac{d\widetilde{\Upsilon}_2}{dz}-\left(\frac{d\widetilde{
\Upsilon}_2}{dz}\right)_{appr}\\
\Delta _1(z)&=&-\Delta _2(z)=-3\,{\frac {{z}^{4}{R}^{3}
 \left( {z_0}-z \right) ^{2}}{ \left( 1+{R}^{6} \left( {z_0}-z
 \right) ^{6} \right) \pi }}+3\,{\frac {{{z_0}}^{4}{R}^{3} \left( z
-{z_0} \right) ^{2}}{ \left( 1+{R}^{6} \left( z-{z_0} \right) ^{ 6}
\right) \pi }}\label{V1m} \eea
  are of order $\gtrsim 10^{-3}$  fm$^3$ only in the interval  $z \in
[z_0^\prime,z_0^{\prime\prime}],  z_0^\prime=4.293\, $
fm,
$z_0^{\prime\prime}=4.307 \,$fm.

Indeed,
 in our consideration (spread case) the
largest value of $z_a$ is $4.260706906$ fm and the  smallest
value of $z_b$ is $4.340400579$ fm. At the points
$z_0^\prime=4.260706906\, $fm, $z_0^{\prime\prime}=4.340400579$ fm
the quantity $\Delta_1$
is less then $\leq 5\cdot10^{-6}$ fm$^3$.

 At the points $z_0^\prime=0.6948439783\, $ fm,
$z_0^{\prime\prime}=1018.393720$ fm the quantity $\Delta_1$
is less then $\leq 2\cdot10^{-12}$ fm$^3$.\\

\begin{figure}[ht] \centering
 \includegraphics[height=5cm]{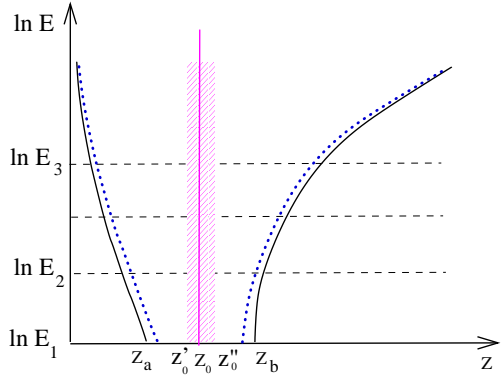}$\,\,\,\,\,$
\caption{
(color on-line)  The schematic
plots of locations of roots (solid black lines) dependent on the
energy (in the logarithmic scale) and the  location  of
differences
$|\displaystyle\frac{d\widetilde{\Upsilon}_i}{dz}-(\displaystyle\frac{d\widetilde{
\Upsilon}_i}{dz})_{appr}|\gtrsim 10^{ -3}$,
$i=1,2$ (the magenta shaded region).   The magenta solid line
shows the location of the wall. The dotted  blue lines show
location of zeros for the charged wall.}\label{root-smoth}
 \end{figure}

The schematic picture of locations of roots and a region there
$|\Delta _i(z)|\gtrsim 10^{-3}$  are presented in
Fig.\ref{root-smoth}. We see that the difference $\Delta_i$ is not
essential in location of the roots and we can use the approximations
(\ref{der-Up-1}) and (\ref{der-Up-2}).

 The regularized version of the the function $\psi$ is
 \be
 \psi_{reg}=\psi_a (z)\Gamma_1+\psi_b(z)\Gamma_2.\ee

Now one has  to put  conditions (\ref{boundary-d}) on the
regularized functions \bea \label{tilde-z}
&&\frac{z_a}{2L}\left.\frac{d}{dz}\psi_{reg}\right|_{z=\tilde{z}_a}=1\\
&&\frac{z_b}{2L}\left.\frac{d}{dz}\psi_{reg}\right|_{z=\tilde{z}_b}=-1\eea
and find $\tilde{z}_a$ and  $\tilde{z}_a$ from these conditions.
However it is difficult to perform these calculations.
 Instead of finding  $\tilde{z}_a$  from
condition (\ref{tilde-z}) we propose to use such regularization
that does not change  $z_a$ found from formal conditions
(\ref{boundary-d}). We can check that the formal $z_a$ in fact
solves also the regularized condition if the regularization is
smooth enough. So, we take $z_a$ and substitute it in the LHS of
 regularized condition (\ref{tilde-z}). We define

$$F_{a,
reg}~\Bigg|_{z=z_a}=\frac{z_a}{2L}\left(\frac{d\psi_a}{dz}\Gamma_1+\frac{d\psi_b}{dz}\Gamma_2\right)~\Bigg|_{z=z_a}=1+\delta_1,$$
$$F_{b,reg}~\Bigg|_{z=z_b}=\frac{z_b}{2L}\left(\frac{d\psi_a}{dz}\Gamma_1+\frac{d\psi_b}{dz}\Gamma_2\right)~\Bigg|_{z=z_b}=-1+\delta_2.$$

We can calculate  $F_{a, reg}$. The deviation of  $F_{a, reg}$
from 1 will
 show
 how the regularization changes
conditions (\ref{boundary-d}). In the following table we present
calculations of $F_{a, reg}$ for the wide range of parameter of
the theory.

We choose the parameter $R$ as minimally needed to make
$\delta_1$ and $\delta_2$ negligible at
 energies $10^{-4}<E<10^2$ TeV. Using the direct numerical
calculations  we choose $R=10^4.$
We perform  numerical calculations at  $R=10^4$ and get the
following table:\\
$$\,$$

\begin{tabular}{|c|c|c|c|c|c|c|}
  \hline
  $E$,\, $TeV$ & $Q$,\, $fm^{1/2}$& $z_a$,\, $fm$ & $z_b$,\,$fm$,& $F_a$ & $F_b$ \\
  \hline &&&&&\\
$118.2$ & $0$ & $0.04399350434$& $4.015208900\cdot10^6$& $1.00000$&$-1.00000$ \\&&&&&\\
  $30$ & $0$ & $0.06948439782$&$1.019088495\cdot10^6$&$1.00000$&$-1.00000$\\&&&&&\\
   $0.03$& $0$ &$0.6948439783$&$1018.393720$&$1.00000$& $-1.00000$\\&&&&&\\
   $0.00025$&$0$&$4.260706906$&$4.340400579$& $0.99999$& $-0.99999$\\&&&&&\\
 \hline
\end{tabular}\\\\

Thus, from the table evidently  $F_a\approx1,$ $F_b\approx-1.$

As has been mentioned above, strictly speaking one may not consider infinite surface as a trapped surface of
 any kind. Nevertheless it is possible to assume that transversal size of colliding objects is finite but very large,
 and therefore boundary conditions do not affect the process of gravitational interactions of inner parts of sources.
 If we are interested only in the specific area of the formed trapped surface in respect to the unit of shock wave area,
we may define it as
\be {\cal A}\approx\lim\limits_{L\rightarrow \infty}\frac{A_{trap}(L)}{A_{source}(L)},\ee
and the approximate equality takes place due to negligibility of boundary effects.
As often happens,  we can  get answers for finite physical systems performing calculations for infinite
non-physical objects.

\section{Holographic QGP  phase diagram  for the central heavy-ions
collisions}

In this section we construct the  phase diagram
for TS formed in the central collision of two identical point-like charged
shock waves \cite{ABJ}.

The profile of point-like charged shock waves in AdS   is given by (\ref{F})
with (\ref{Fp}) and (\ref{FQ}).
Existence of the trapped surface in  the central collision of two point-like charged
shock waves  means the existence of a real solution, $q_0$,   to the following
equation (see \cite{ABJ} for details)
\be
\label{qc-eq-m}
F^\prime (q_0)-\frac{2}{1+2q_0}F(q_0)+\frac{2L}{\sqrt{q_0(1+q_0)}}=0
\ee
The left hand side of \eqref{qc-eq-m} can be written as
\be
\label{d-na}
{\cal F}(L,E,\bar{Q}^2,q)=
\frac{{\cal N}(L,\bar{M},\bar{Q}^2,q)}{{\cal D}(a,q)}.
\ee
 The  numerator ${\cal N}(L,E,\bar{Q}^2,q)$
contains just one  term
with dependence on $\bar{Q}^2$. This dependence  is linear
with a positive coefficient
\be
{\cal N}(a,\bar{M},\bar{Q}^2,q)={\cal N}(a,\bar{M},q)+
15 \frac{ \pi}{a}\,\bar{Q}^2.
\ee
 The denominator in (\ref{d-na})  does not take infinite values.
 To find solutions to  (\ref{qc-eq-m})
for the shape function given by (\ref{F}) we can draw the function
\bea
\nonumber
-{\cal N}(a,\bar{M},q)&\equiv& -(512a^3 q^5+1280 a^3 q^4-
96 \bar{M}\pi a q^2+1024 a^3 q^3-96\bar{M} \pi a q\\&+&256 a^3 q^2),
\eea
and  see where this function  can be equal
 to a given value
$
15 \bar{Q}^2\frac{ \pi}{a}.$

In order to find the maximal allowed $\bar Q^2$ at which solution to \eqref{qc-eq-m} still exists
we find the maximum of function ${\cal N}$
for fixed energy,
\be
\frac{d{\cal N}(a,\bar{M},q)}{dq}|_{q=q_{max}}=0\ee
and the value
 $$\frac{a}{15\pi }{\cal N}(a,\bar{M},q)|_{q=q_{max}}$$
defines $\bar{Q}_{
max}^2$.

Let us remind that we are
 working in physical units and we use the following  notations
 (\ref{Qbar}) and (\ref{Mbar}): $\bar{M}=\displaystyle\frac{4G_5E}{3\pi}$
 and $\bar Q^2=
\displaystyle\frac{4G_5 Q_n^2}{3\pi}$.\\

\begin{figure}[ht] \centering
 \includegraphics[height=7cm]{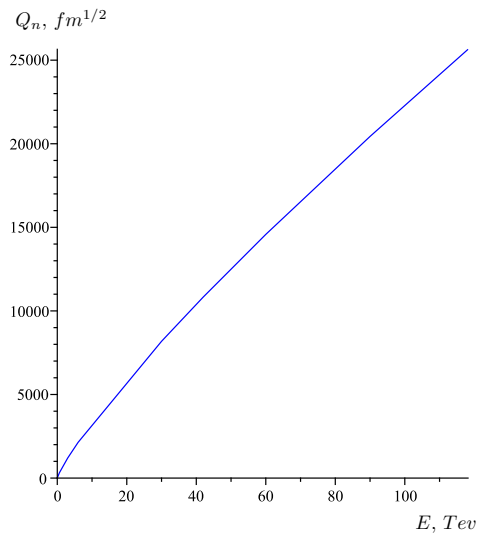}
\caption{The allowed zone for the trapped surface formation is under
the line on the diagram. The plot has been constructed by using
formulas from \cite{ABJ}. } \label{F-ABJ}\end{figure}

Results of calculations are presented in Fig. \ref{F-ABJ}.

\newpage

To estimate corrections to  GYP multiplicity  due to  non-zero
chemical potential, we use  formula (3.17) from \cite{ABJ}. In
notations admitted in this paper, (\ref{Fp}) and (\ref{FQ}), the
formula has the form \bea A_{AdS_5} &\approx&4\pi L^3
\left(\frac{G_5E}{L^2}\right)^{\frac{2}{3}}
\left(1-\frac{1}{24}\left(1+\frac{5Q_n^2}{EL^2}\right)
\left(\frac{2\sqrt{2}L^2}{G_5E}\right)^{\frac{3}{2}}\right)\eea

\begin{figure}[ht] \centering
 \includegraphics[height=6cm]{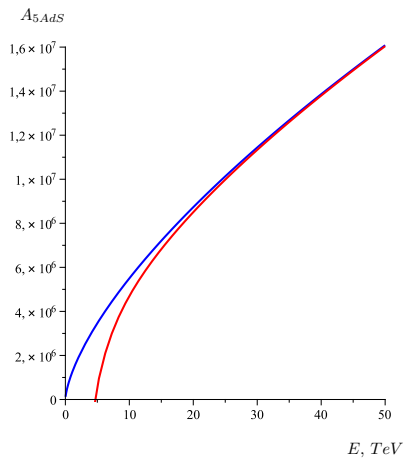} \caption{ The function $A_{AdS_5}$, at  $Q_n=0,$ (blue line)
 and  $Q_n=2\cdot10^6 fm ^{1/2}$ (red line).
}\label{S-QQ}\end{figure}

In Fig.\ref{S-QQ}  we show the entropy, $A_{AdS_5}$, for $Q_n=0$
and $Q_n\neq 0$. The blue line represents
 $Q_n=0$.  The red line represents
 $Q_n=2\cdot 10^6$. We see that the deviation form the GYP multiplicity is essential for small energies
 and is almost  neglectful  for large energies.

\section{Holographic QGP  phase diagram  in the wall-wall dual model of heavy-ions
collisions} \subsection{Charged wall as a dual model for a heavy-ion
with non-zero chemical potential}
Let us note that
the form of the $J^{WP}_{uu}$ in (\ref{EinsteinwithoutQ}) can be obtained
         by spreading out the  energy-momentum tensor of an ultrarelativistic point,
         i.e
         $J_{uu}$ in the form (\ref{rho}) with $\rho(q)$ given by eq.( \ref{psw-rho}),
 over the
transversal surface.

The Einstein equation for
    the  charged wall (membrane)   has  the form
          \begin{equation}\label{withQQ}
             (\partial^2_z-\frac{3}{z}\partial_z)
             \phi(z)=-16\pi
             G_5\left(J^{WP}_{uu}+J^{WQ}_{uu}(Q,z)\right).
             \end{equation}
             where $J^{WP}_{uu}$ is given by (\ref{EinsteinwithoutQ})
             and we suppose that $J^{WQ}_{uu}(Q,z)$ can be obtained in the similar way
by spreading  the  energy-momentum tenzor of the ultrarelativistic charged point
         $T^{pQ}_{uu}$ over the
transversal surface. In the previous calculations:

\be
\label{T-J-int-m}
J^{WQ}_{uu}=\frac{\int _{\cal M}J^{pQ}_{uu}{\cal D}x_{\perp}}{
\int _{\cal M}{\cal D}x_{\perp}}
\ee
here  the subscript "pQ" means the electromagnetic part of the energy momentum tensor
of the charged point particle and "${\cal D}x_{\perp}"$
means that we integrate over the induced metrics on the orthogonal surface ${\cal M}$.

For this purpose we take
\be
 J^{pQ}_{uu}(z,z_0)=\frac{L}{z}\rho^{pQ}\ee
where $\rho^{pQ}$ is given by (\ref{rho-charge}), and according to our prescription
(\ref{T-J-int-m})
we integrate over all transversal coordinates
\bea\label{withQ}\quad J^{PQ,II}_{uu}&=&
\frac{\frac{L}{z}\int_{0}^{\infty}\rho^{pQ}(q)
\frac{L^2}{z_0^2}
   \frac1{2}dr^2}{\int_{0}^{\infty}
  \frac{L^2}{z_0^2} r dr}
\eea

The result is
\be
J^{pQ}_{uu}={\cal X}{\cal J}\label{J}\ee
where
\bea
\label{Jcal}
{\cal J}&=&\frac{64}{3}zz_0\left( 1-\frac{z_0^{6}-3\,{z}^{2}z_0^{4}-3\,{z}^{4}z_0^{2}+{z}^{6}
}{|z_0^{2}-{z}^{2}|^3}\right)
 \\
 {\cal X}&=&\frac{5}{256}\frac{Q_n^2}{\pi L^6}=\frac{5}{256}\frac{{Q}^2}{L^6}\,\,\,\eea

 We see divergency at $z=z_0$, as it
should be for the energy-momentum tensor of a charged plane. We introduce
regularization by adding the $\epsilon$ factor in the denominator.

\subsection{Charged
wall-on-wall collision as a dual model for heavy-ions collisions with non-zero
chemical potential}
To find the TS formation condition in the wall-wall collision one has to
solve  Einstein equation
          \bea\label{withQ-m}
             (\partial^2_z-\frac{3}{z}\partial_z)
             \phi(z)&=&-16\pi
             G_5\left(J^{pW}_{uu}(z)+J^{QW}_{uu}(Q,z)\right),\\
             \\
             J^{pW}_{uu}(z)&=&\frac{E}{L^2}\frac{z_0^3}{L^3}\delta(z-z_0),\\
             J^{QW}_{uu}(Q,z)&=&
             \frac{128{\cal X}}{3}zz_0{\frac {\,{z}^{4}\left( -{z}
^{2}+3\,{{z_0}}^{2} \right)\theta(z_0-z)+ z_0^{4}\left(-3{z}^{2}+z_0^{2}\right)\theta(z-z_0)}{\left(
-{z}^{2}+z_0^{2}+{\epsilon}^{2} \right) ^{3}}}\nonumber\\
             \eea
with the following boundary conditions
\bea
   \bf {1})&\,\,\,& \phi(z_a)=\phi(z_b)=0,\phi_a(z_0)=\phi_b(z_0)
   \label{1-st-cond}\\
   \bf {2})&\,\,\,& \left(\psi^{\prime}(z_a)\displaystyle\frac{z_a}{L}\right)=2,\,\,\,\,\,
   \left(\psi^{\prime}(z_b)\displaystyle\frac{z_b}{L}\right)=-2,
   \label{2-nd-cond}
\eea
where $z_a$ and $z_b$ are the boundaries of the TS and  $\psi$ is related to
\be
\phi(z)=\frac{z}{L}\psi.
\ee

    We search for a solution to the Einstein equation  with a charged source in the form
    of the sum of the "neutral" solution and a correction proportional to $Q^2$
 \be
 \phi=\phi_n+\phi_q
 \ee
 here $\phi_n$ denotes the solution of the neutral case.

As in the neutral case it is convenient to  consider domains $z<z_0,$ $z>z_0$ separately
  \begin{equation}\label{complitesolution}
\phi_q=\left\{
\begin{array}{cc}
  \phi_{qz_0>z},\,\,\,\,\, z_0>z;\\
  \phi_{qz>z_0},\,\,\,\,\, z>z_0\\
\end{array}
\right.
\end{equation}
and we have
\bea\label{eqz+z}
  &&(\partial^2_z-\frac{3}{z}\partial_z)
    \phi_{qz_0>z}=-16\pi  G_5{\cal X}\frac{128}{3}zz_0{\frac {\,{z}^{4}\left( -{z}
^{2}+3\,{{z_0}}^{2} \right) }{\left(
-{z}^{2}+z_0^{2}+{\epsilon}^{2} \right) ^{3}}},\, z_0>z;\\
&&(\partial^2_z-\frac{3}{z}\partial_z)
    \phi_{qz>z_0}=-16\pi G_5 {\cal X}\frac{128}{3}zz_0
    \frac {z_0^{4}\left(-3{z}^{2}+z_0^{2}\right)}{\left(
-{z}^{2}+z_0^{2}-\epsilon^{2}\right)^{3}}\,,
    z>z_0.\label{eqz-z_0}
\eea

Solutions to (\ref{eqz+z}) and
(\ref{eqz-z_0}) can be presented as :
\begin{eqnarray}
\psi_{qz_0>z}&=&{z}^{3}{C_1}+{\frac {{C_2}}{z}}-{\frac {NL{z_0}\,{z}^{3}}{4\left(-{z}^{2}+{{z_0}}^{2}+\epsilon^2 \right)}},\quad z_0>z,\label{sa}\\
  \psi_{qz>z_0}&=&{\frac
  {{C_3}}{z}}+{z}^{3}{C_4}+{\frac {NL{{z_0}}^{5}}{4z\left(-{z}^{2}+{{z_0}}^{2}+{\epsilon}^{2}
 \right)}},\quad z>z_0\label{sb}
\end{eqnarray}
Here $N=\frac{40}{3}\frac{\pi G_5
Q^2}{L^6}$
The first two terms
in (\ref{sa}) and (\ref{sb})  are solution to the Lin and Shuryak
equation (55) in \cite{Shuryak09}. If one assumes that they satisfy
condition {\bf {1}}, i.e. $\psi_n(z_a)=\psi_n(z_b)=0,$
$\psi_{na}(z_0)=\psi_{nb}(z_0)$,  one gets \cite{Shuryak11}:
\begin{eqnarray}
 \label{solution 11}\Psi_n&=&\left\{%
\begin{array}{cc}
  \psi_{na}=C\left(\displaystyle\frac{z^3}{z_a^3}-\displaystyle\frac{z_a}{z}\right),\, C=-\displaystyle\frac{4\pi G_5 E}{L^4}\frac{\left(\displaystyle\frac{z_0^4}{z_b^4}-1\right)z_b}{\displaystyle\frac{z_b^4-z_a^4}{z_a^3z_b^3}},\, z<z_0
  \\\\
  \psi_{nb}= D\left(\displaystyle\frac{z^3}{z_b^3}-\displaystyle\frac{z_b}{z}\right),\, D=-\displaystyle\frac{4\pi G_5 E}{L^4}\frac{\left(\displaystyle\frac{z_0^4}{z_a^4}-1\right)z_a}{\displaystyle\frac{z_b^4-z_a^4}{z_a^3z_b^3}},\,z_0<z
\end{array}%
\right.
\end{eqnarray}
In the neutral case one find $z_{a}$ and $z_b$ from the 2-nd condition
$\left(\psi_{na}^{\prime}(z_a)\frac{z_a}{L}\right)=2,
\,\left(\psi_{nb}^{\prime}(z_b)\displaystyle\frac{z_b}{L}\right)=-2,$
here $z_a$ and $z_b$ are the boundaries of the TS.

As to (\ref{sa}) and (\ref{sb}), choosing
\bea
  C_1&=& \frac{NLz_0}{4(z_a^2-z_0^2)},\quad C_2=0,\\
  C_3&=&\frac{NLz_0^5}{4(z_b^2-z_0^2)},\quad C_4=0,
\eea
 we obtain

\begin{equation}\label{chargpart}
    \left\{%
\begin{array}{cc}
  \psi_{aq}=-\displaystyle\frac{NLz_0z^3}{4}\displaystyle\frac{-z_a^2+z^2-\epsilon^2}{(-z^2+z_0^2+\epsilon^2)(-z_a^2+z_0^2)}, & z<z_0
  \\\\
  \psi_{bq}=\displaystyle\frac{NLz_0^5}{4z}\displaystyle\frac{-z_b^2+z^2+\epsilon^2}{(-z^2+z_0^2-\epsilon^2)(-z_b^2+z_0^2)}, & z_0<z\\
\end{array}%
\right.
\end{equation}

Note that for the constructed solution  the
condition $\psi(z_a)=\psi(z_b)=0$ is satisfied  automatically.

The second requirement  (\ref{2-nd-cond}) gives

\begin{eqnarray}
 &&-\frac{8\pi G_5 \,E\left({z_0^4-z_b^4}\right)z_a^3}{L^5(z_b^4-z_a^4)}-\displaystyle\frac{N}{4}\frac{z_0z_a^5}{(-z_a^2+z_0^2)^2}=1,\label{A_1}\\
  &&-\frac{8\pi G_5 \,E\left({z_0^4-z_a^4}\right)z_b^3}{L^5(z_b^4-z_a^4)}+\displaystyle\frac{N}{4}\frac{z_0^5z_b}{(-z_b^2+z_0^2)^2}=-1;\label{B_1}
\end{eqnarray}
 These equations do not have analytical solutions and we
treat them numerically.

Roots of system \eqref{A_1},\eqref{B_1} could not be
found analytically since these equations are
equivalent to polynomial equations on $z_a$ and $z_b$ of a high
degree ($>4$). So we take $z_0=L$ and analyze the following system numerically
\begin{eqnarray}
\label{Trapped-radius1}
 F_a&\equiv&-\frac{8\pi G_5 E \left({z_0^4-z_b^4}\right)z_a^3}{z_0^5(z_b^4-z_a^4)}-\frac{10}{3}
 \frac{\pi G_5Q^2}{z_0^6}\frac{z_0z_a^5}{(-z_a^2+z_0^2)^2}=1,\label{A} \\
 \label{Trapped-radius2}  F_b&\equiv&-\frac{8\pi G_5 E \left({z_0^4-z_a^4}\right)z_b^3}{z_0^5(z_b^4-z_a^4)}+\frac{10}{3}
 \frac{\pi
 G_5Q^2}{z_0^6}\frac{z_0^5z_b}{(-z_b^2+z_0^2)^2}=-1.
\end{eqnarray}

\begin{figure}[ht] \centering
 \includegraphics[height=6cm]{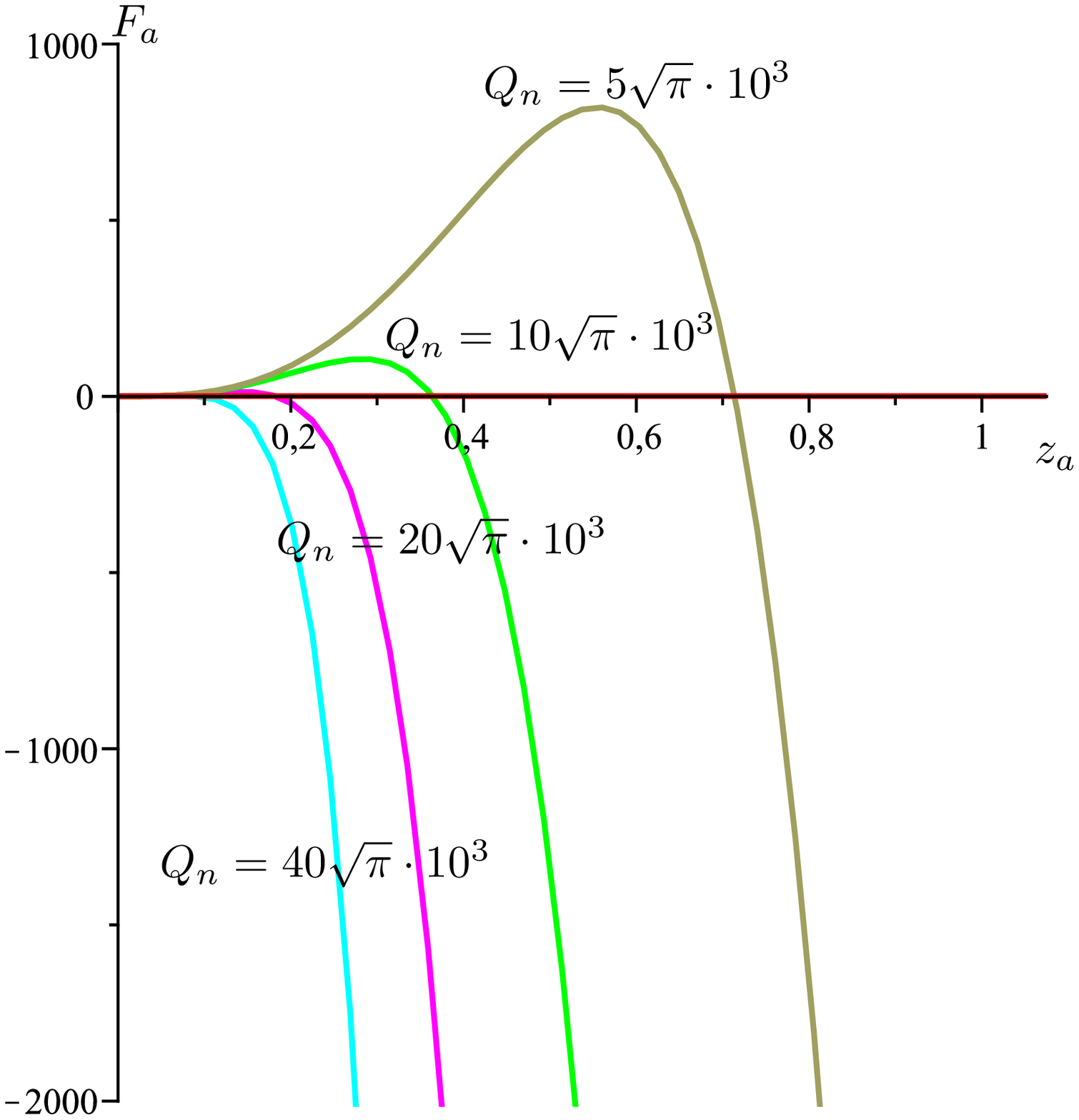}A.\,\,\,\,\,\,\,\,
 \includegraphics[height=6cm]{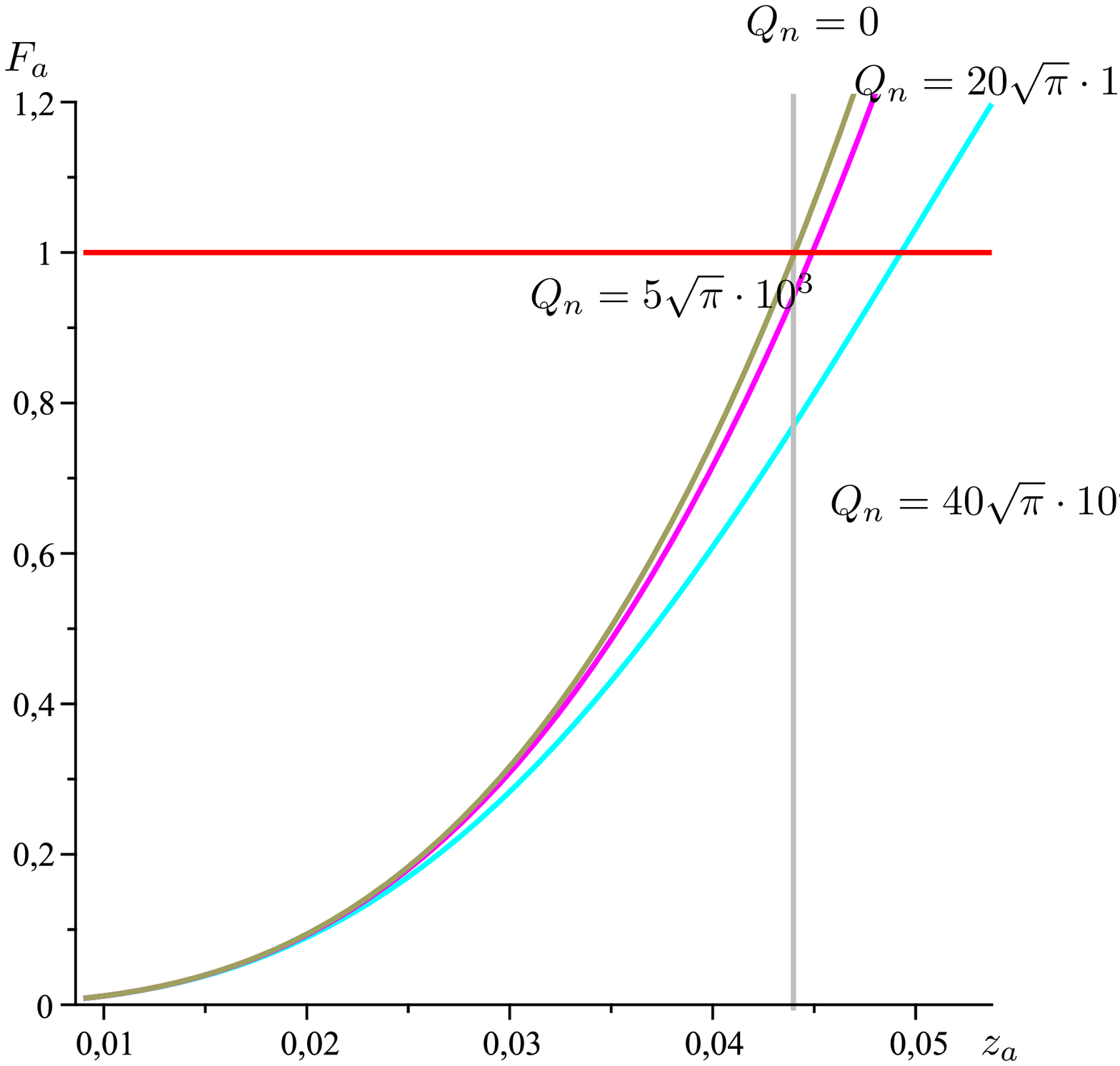}B.
\caption{ The plot of  $F_a(z_a,z_b)$
as a function of $z_a$ for fixed $z_b$ near the root $z_a=z_a(E)$ at $E=118.2$ TeV.
 Figure (B) zooms in the region
of small $F_a$ and small $z_a$.} \label{Fa}\end{figure}

\begin{figure}[ht] \centering
 \includegraphics[height=6cm]{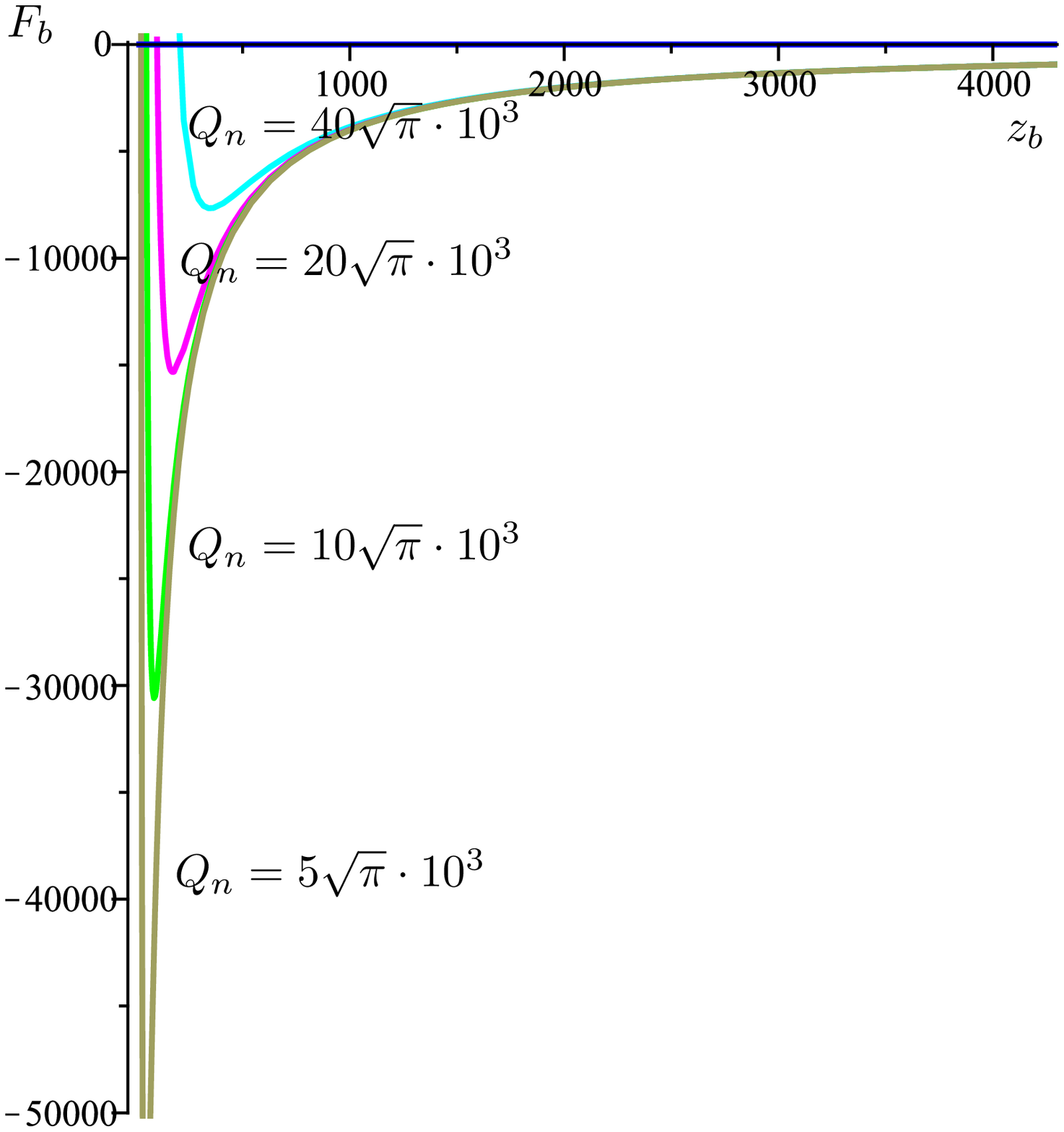}A.\,\,\,\,\,\,\,\,
 \includegraphics[height=6cm]{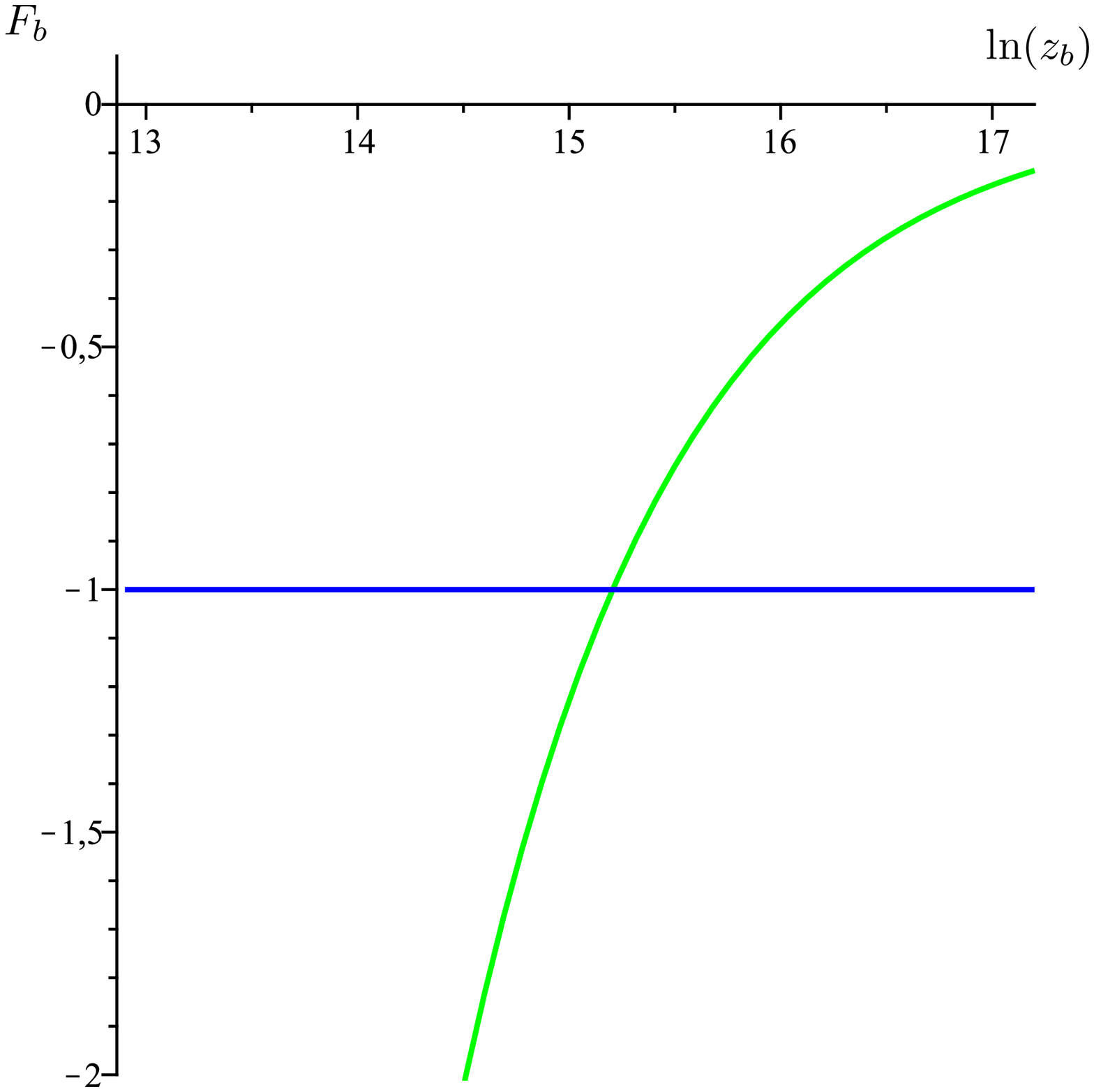}B.
\caption{
 The plot of  $F_b(z_a,z_b)$
as a function of $z_b$ for fixed $z_a$ near the root $z_b(E)$ at $E=118.2$ TeV.
 Figure B zooms in the region
of small negative $F_b$ and presents  $z_b$ in the logarithmic scale. } \label{Fb}\end{figure}

To show the movement of roots of equations (\ref{Trapped-radius1})
and (\ref{Trapped-radius2}) we
 suppose that  $z_b$ for given $Q$ is already known  and
represent function $F_a(z_a,z_b)$ as function of $z_a$ in Fig. \ref{Fa}.
In the similar way, supposing that $z_a$ is already known we represent
function $F_b(z_a,z_b)$ as function of  $z_b$ in Fig. \ref{Fb}.

\newpage

\begin{figure}[ht] \centering
 \includegraphics[height=7cm]{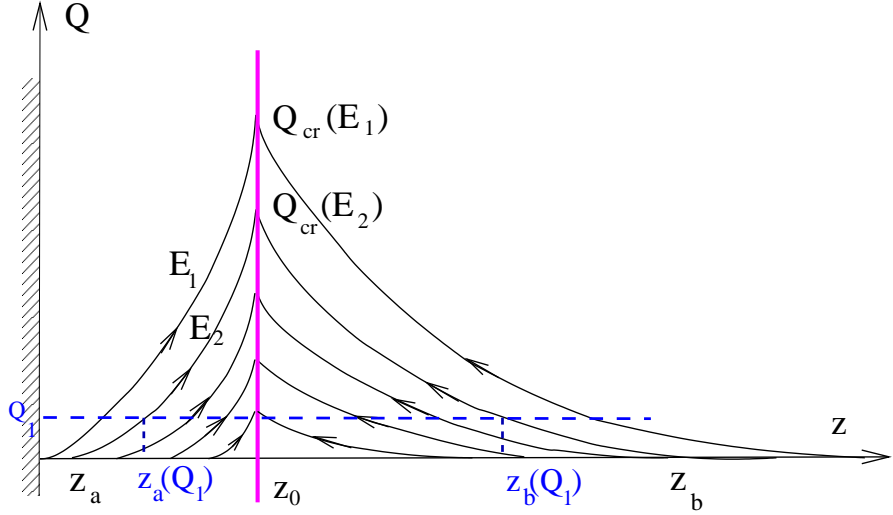}
\caption{ The schematic picture of  charge flows
The magenta solid line shows the position of the wall.
 We see that the positions of points $z_a(Q)$ and $z_b(Q)$ move to the point $z=z_0$
 when we increase $Q$.
For $Q\to Q_{cr}(E)$ the segment $[z_a(Q),z_b(Q)]$ shrinks to zero.}
 \label{RFSchem}\end{figure}

\begin{figure}[ht] \centering
 \includegraphics[height=7cm]{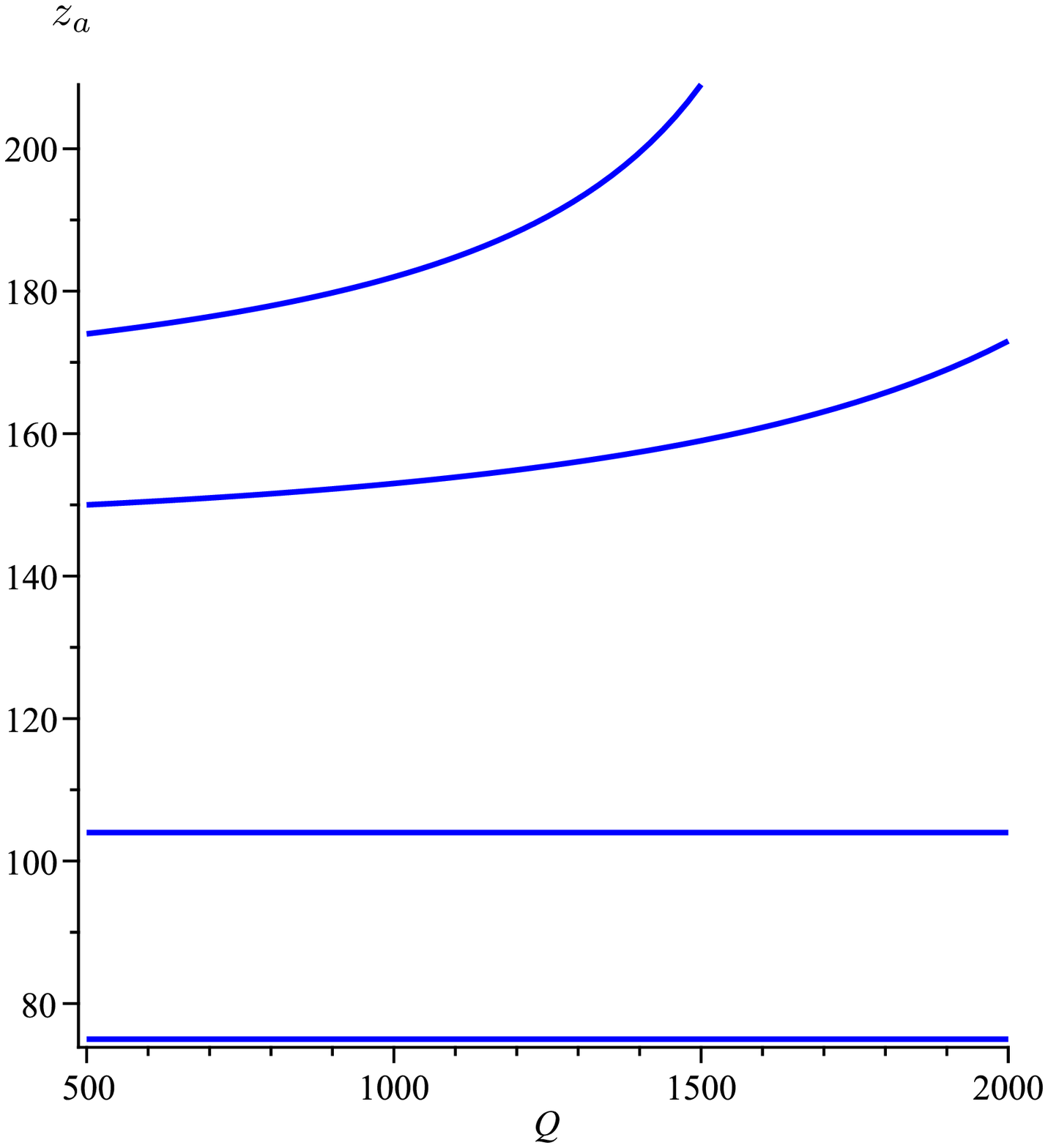}$\,\,\,\,\,\,\,\,\,\,\,\,\,\,\,\,\,$
 \includegraphics[height=7cm]{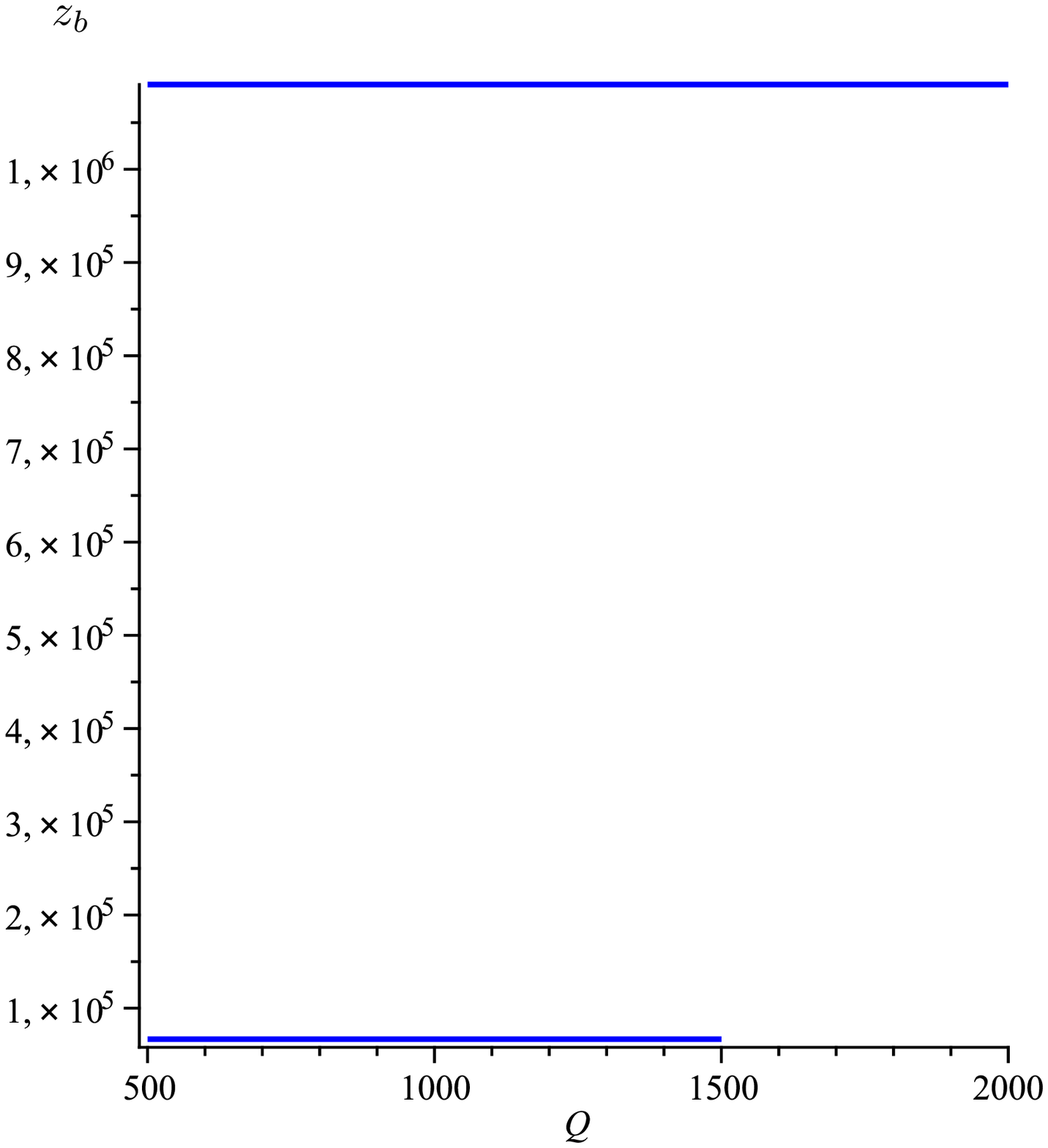}
\\$\,$\caption{
A. The charge flows of the  root $z_a(Q)$ for  $E=1.97\,$TeV, $3 \,$TeV, $9
\,$TeV, and $24\, $TeV. B. The charge flows of the  roots $z_b(Q)$ for $E=1.97
\,$TeV, $3\, $TeV. } \label{Rflow}\end{figure}

In Fig.\ref{RFSchem} we show the charge flows of the roots.
Different  lines correspond to
 different energies. We see that the flows go to $z_0$
and reach the line $z=z_0$ for $Q=Q_{cr}$. In Fig.\ref{Rflow}
we draw the corresponding flow for physical parameters.

\subsection{Comparison of results.}

\begin{figure}[ht] \centering
 \includegraphics[height=5cm]{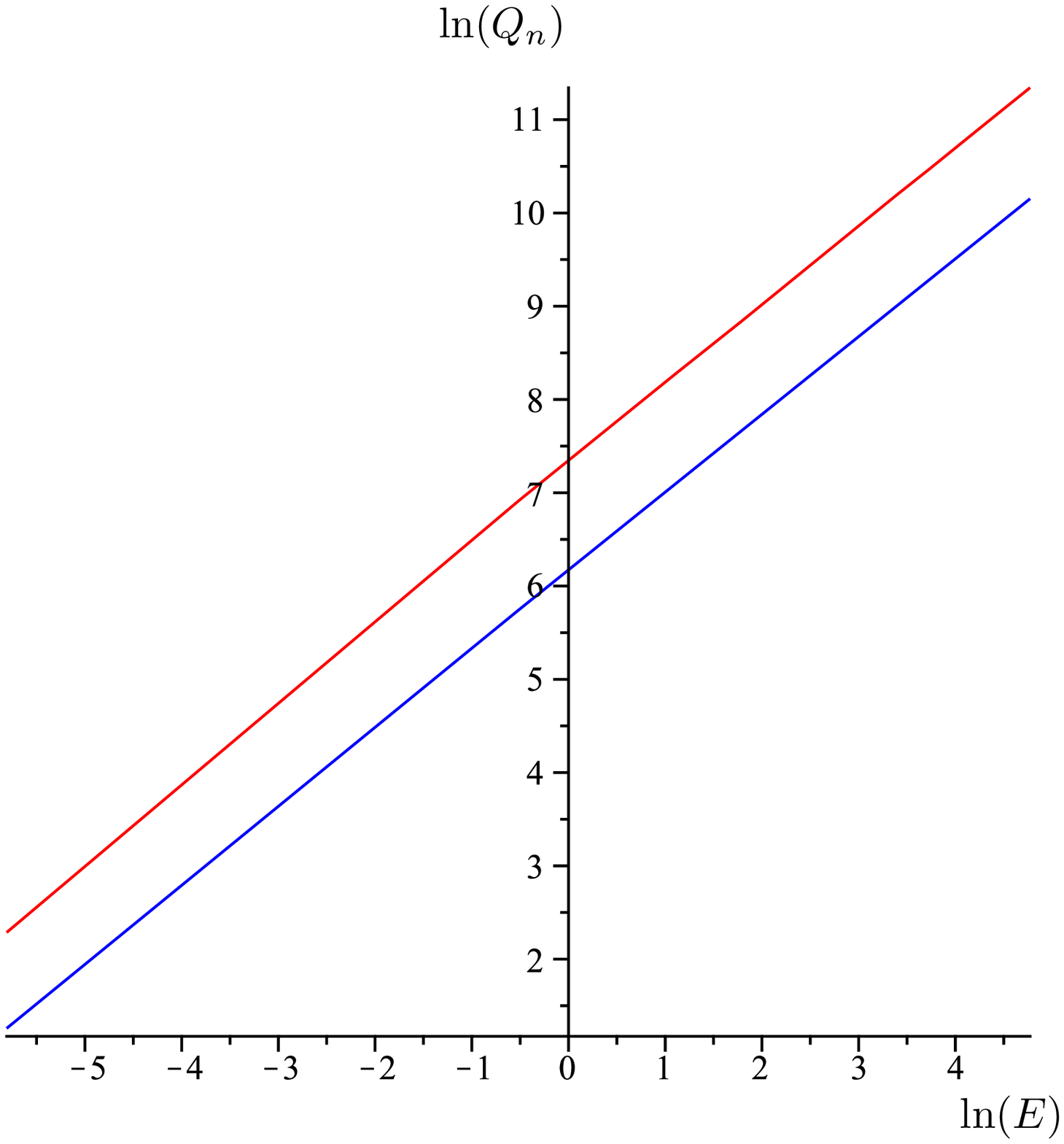}$A.\,\,\,\,\,\,\,\,$
 \includegraphics[height=5cm]{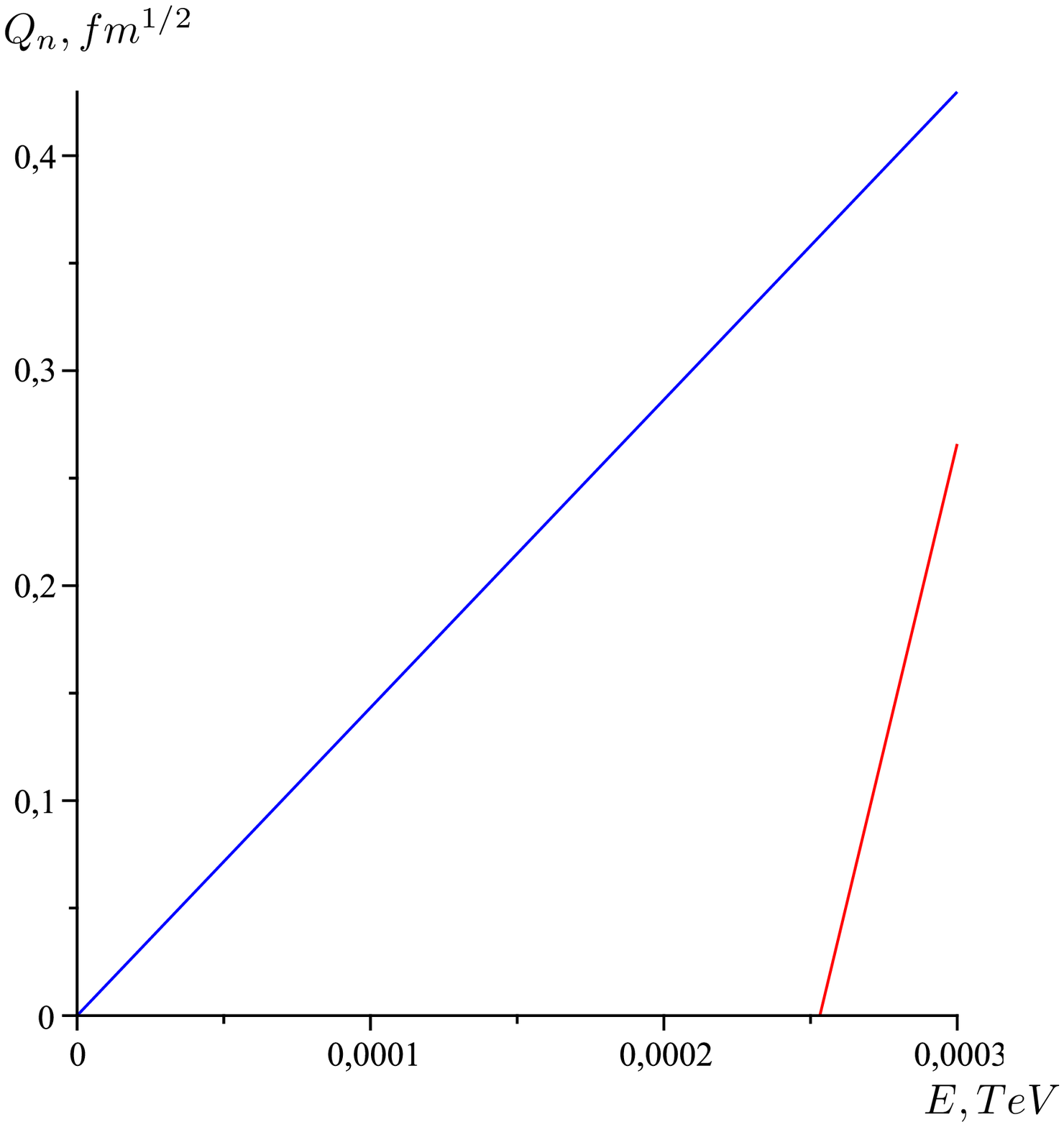}$B$
\caption{A. The phase diagram  the logarithm of  $Q_n$ vs the logarithm of $E$
at large $E$.
 B. The phase diagram $E$ vs $Q_n$ for small $E$ and small $Q_n$.
  The blue lines correspond to the pointlike charge and
 the red lines  to  the spread charge.
 The zones above  the lines are forbidden for  black holes production
 for corresponding $E$ and $Q$.}
 \label{Q-E}\end{figure}

It is interesting to compare the phase diagrams, the energy (temperature) $E$ vs
the charge (chemical potential) $Q$, corresponding to the pointlike charge and
  the spread charge. Results of these calculations are collected in the table below and
  presented in Fig.\ref{Q-E}. We see that this two phase diagram are almost the same.
$$ $$
\begin{tabular}{|c|c|c|c|c|c|c|c|c|c|c|}
  \hline
  $E$ (TeV) & $118.2$&  $60$  & $30$ & $6$ &  $3$  & $0.6$ &  $0.06$ & $0.03$  &  $0.0003$ &  $0.00025$ \\
  \hline
  $Q_{cr,\, point}$   & $25649.6$ & $14577.2$  & $8180.6$  & $2138.7$   & $1199.9$ & $313.3$ & $45.6$ & $25.4$ &$0.43$ &  $0.37$ \\
  $Q_{cr, \,wall}$  &$47500 $ & $27000 $&  $15170 $ & $3950 $ & $2220$  & $570 $ & $80$ & $40 $ & $0.15$ & $0$ \\
  \hline
\end{tabular}

\begin{figure}[ht] \centering
 \includegraphics[height=5cm]{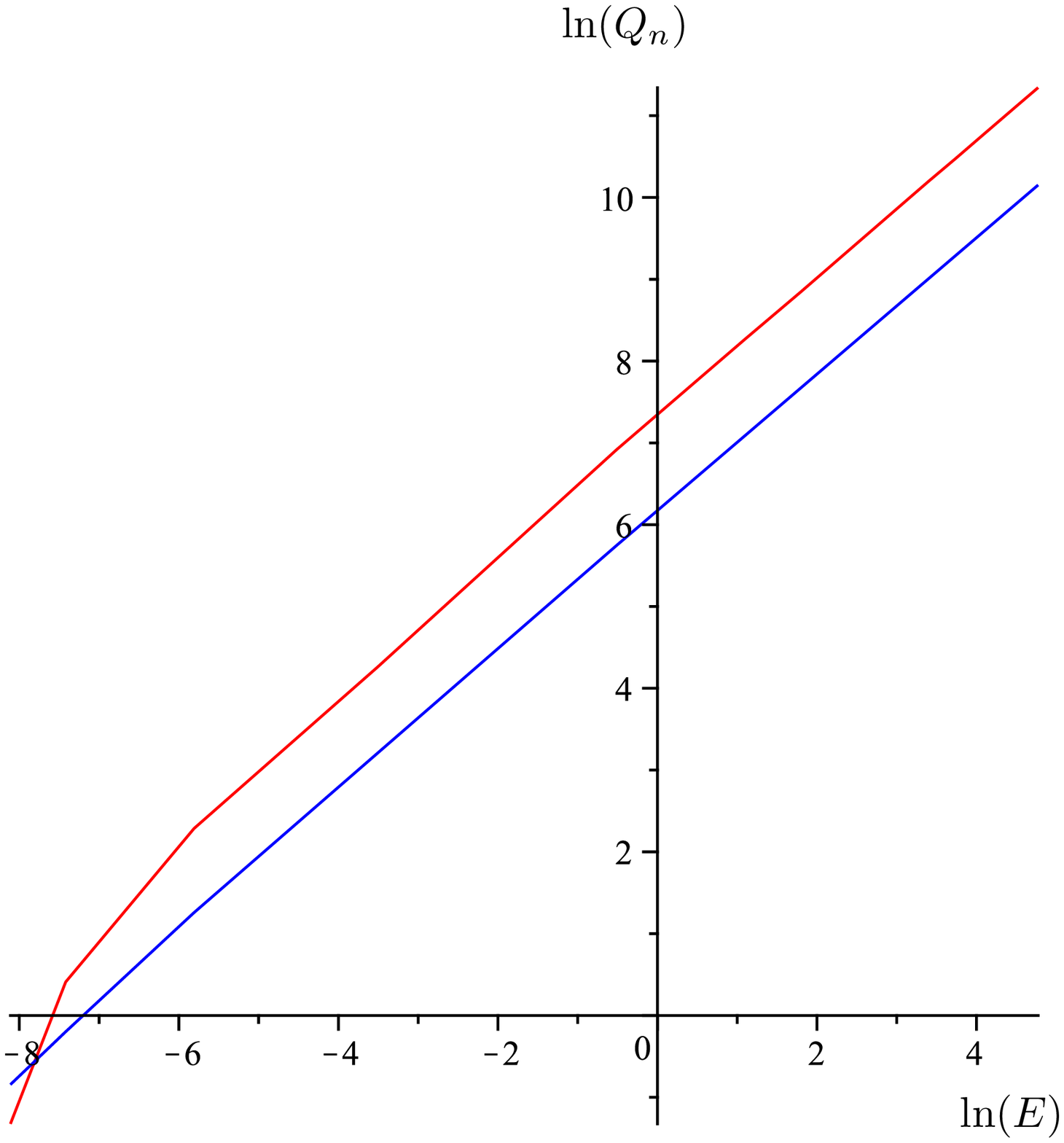}$\,\,\,\,\,\,\,\,$
 \includegraphics[height=5cm]{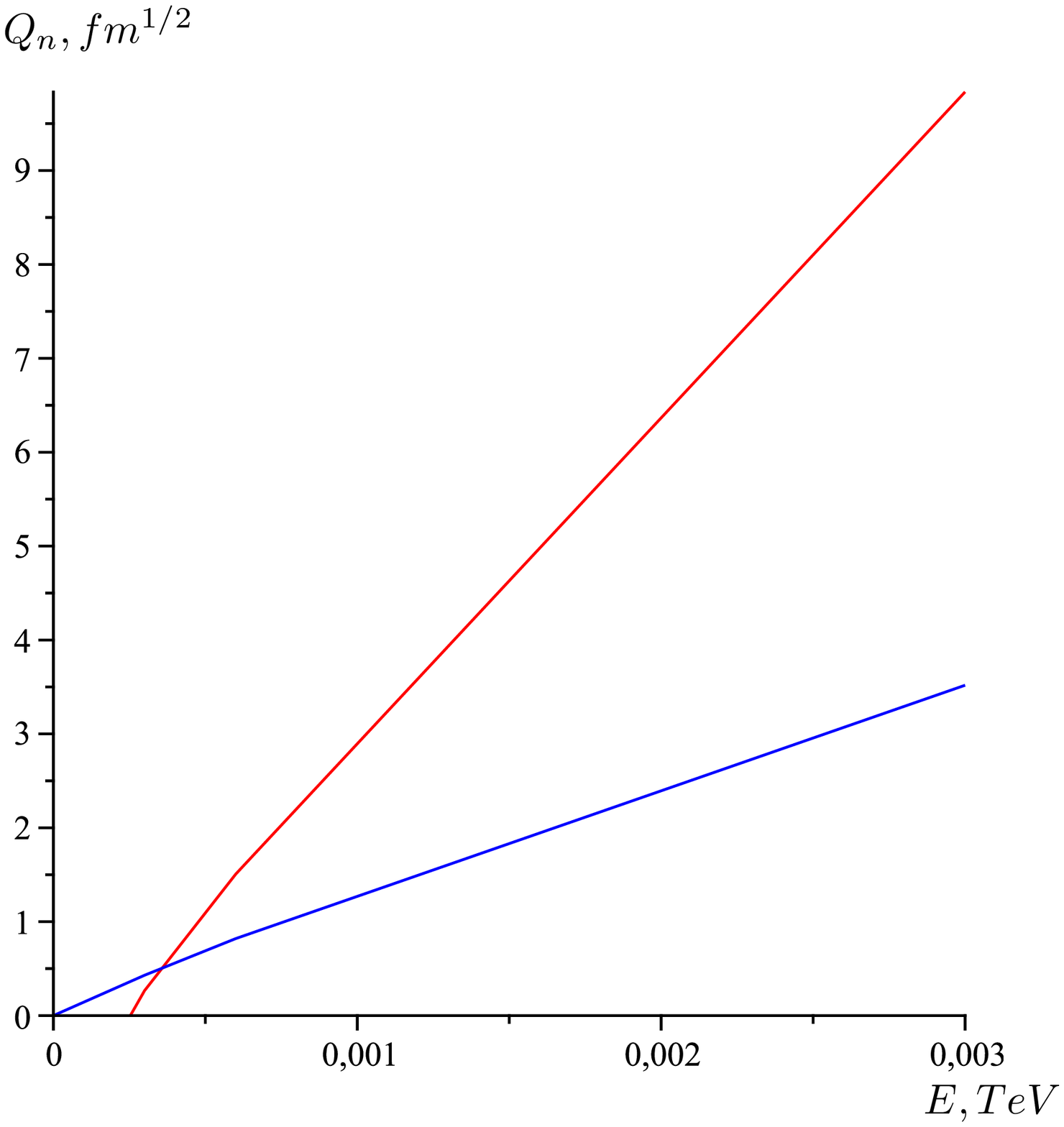}
 \caption{The cross point of two diagrams in logarithmic and natural scales.}
 \label{cross point}\end{figure}

From Fig. \ref{Q-E} it is evident that the two lines, the red and the blue ones,  have
a cross point. We represent the cross point in natural and  logarithmic scales
in Fig. \ref{cross point}.

\subsection{The square trapped surface calculation}

Following \cite{Shuryak09} we calculate entropy lower bound as
"the area of the trapped surface" per an unite square of the
wall\footnote{We put "area" and "trapped surface" in quotation
marks since in the strict notions of the trapped surface it has to
be smooth and compact. In our case it is not smooth and it does
not have finite area, one can only assume this properties after
regularization  } using the formula:
\begin{eqnarray}\label{trapped}
  &&S=\frac{2A}{4G_5}=\frac{\int\sqrt{g}dzd^2x_\perp}{2G_5},\\
  \label{trapped-S}
  &&s\equiv\frac{S}{\int d^2x_\perp}=\frac{L^3}{4G_5}\left(\frac{1}{z_a^2}-\frac{1}{z_b^2}\right).
  \end{eqnarray}

 In the absence of
 transverse dependence one ignores $x^2_{\perp}$ in (\ref{trapped}).
(\ref{trapped-S})  measures entropy per transverse area.

\begin{figure}[ht] \centering
 \includegraphics[height=5cm]{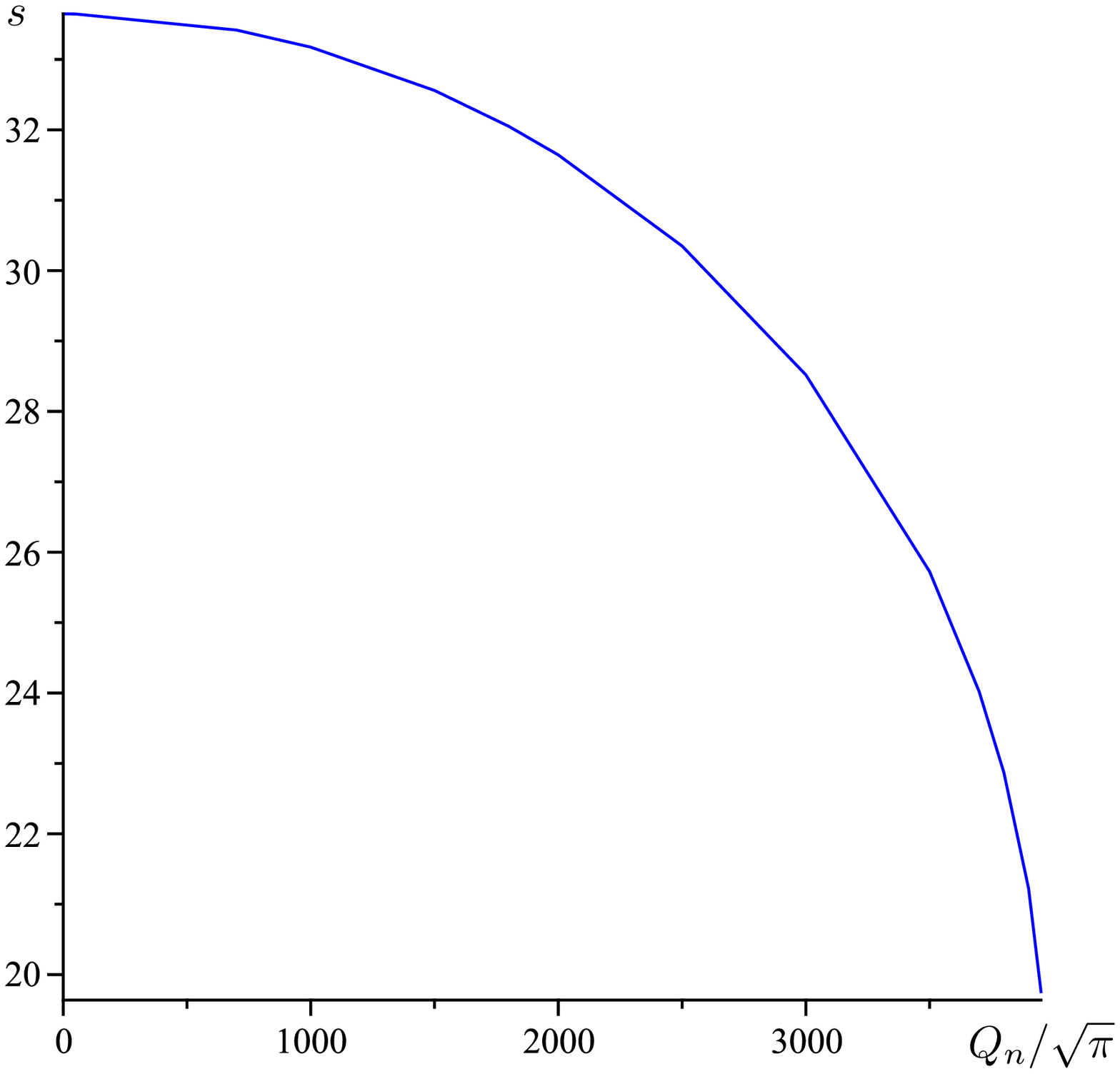}\,\,\,\,\,\,\,\,
 \includegraphics[height=5cm]{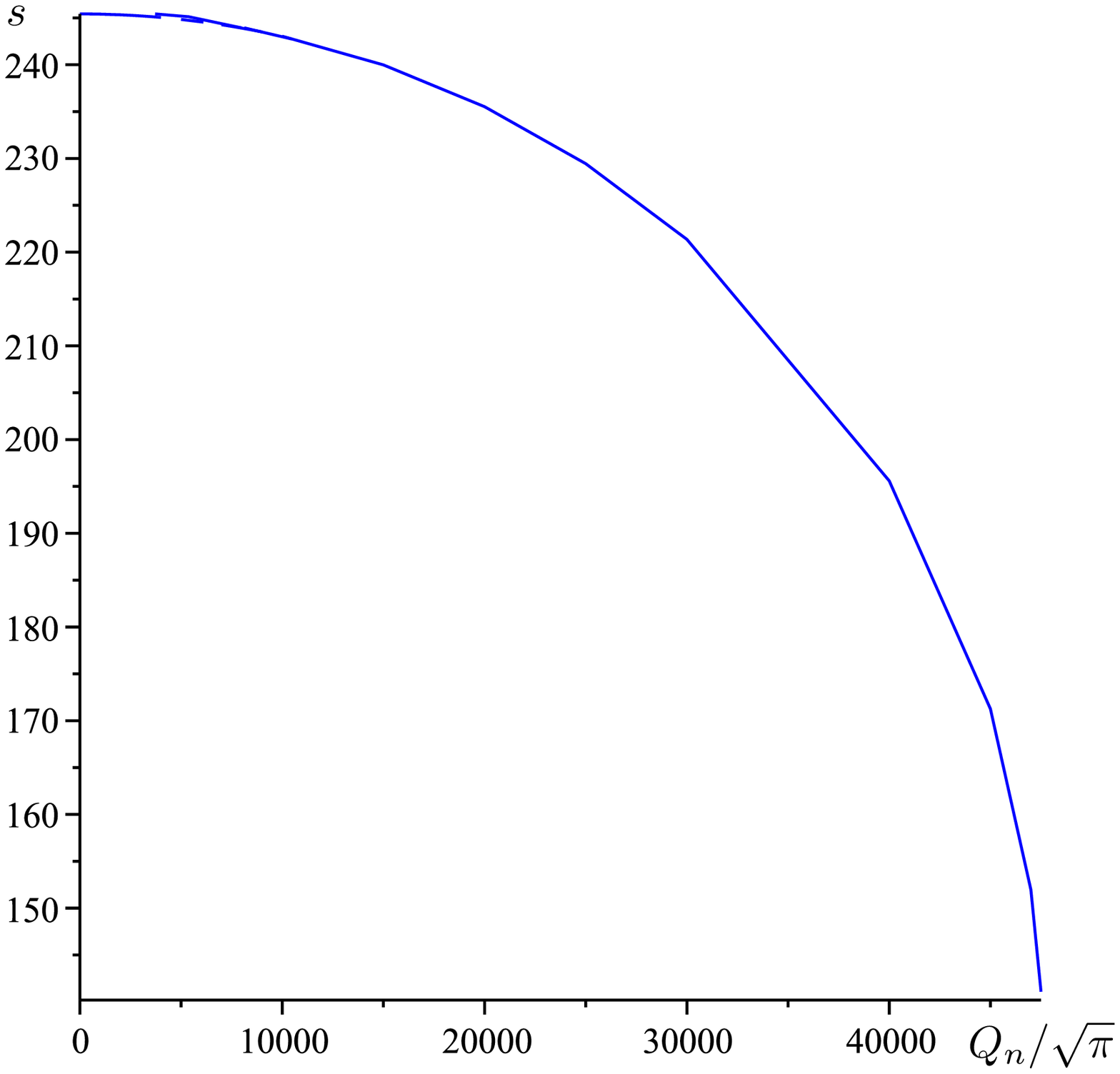}
\caption{The dynamics of the trapped surface area $s(Q_n/\sqrt{\pi})$
at $E=6\, TeV,$ $E=118.2\,TeV$. } \label{Surface}\end{figure}

The
trapped surface decreases with growth of a charge. The
corresponding graphical representations are in  Fig.
\ref{Surface}.

\begin{figure}[ht] \centering
 \includegraphics[height=7cm]{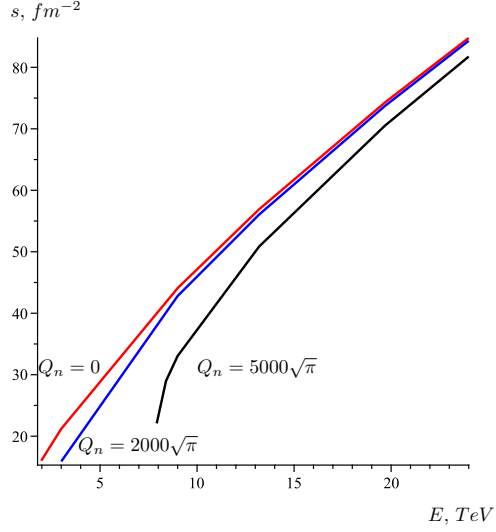} \caption{The red
line corresponds to the case $Q_n=0\, fm^{1/2},$ the blue to the
case $Q_n=2000\sqrt{\pi}\, fm^{1/2},$ the black to the case
$Q_n=5000\sqrt{\pi}\,fm^{1/2}.$}\label{S(E)Qi}\end{figure}

In Fig.\ref{S(E)Qi} we show the entropy per volume given by  (\ref{trapped-S})
as function of energy for different $Q$. This plot is similar to the plot presented in Fig. \ref{S-QQ}.
   We see that the influence of the chemical potential on  the multiplicity is essential for small energies
 and is almost  neglectful  for large energies.

\subsection{ Remarks about the regularization }

 The regularized version of the the function $\psi$ is
 \be
 \psi_{reg}=\psi_a (z)\Gamma_1+\psi_b(z) \Gamma_2\ee
 where $\psi_a (z)$ and $\psi_b(z)$ define the function  $\psi$
 without regularization,
 \be
 \psi=\psi_a (z)\Theta (z_0-z)+\psi_b(z) \Theta(z-z_0)\ee

$$\psi_a(z)=- \frac{4\,G\pi \,E\left( {\frac {z_0^{4}}{{{z_b}}^{4}}}-1
 \right) {{z_b}}^{4}{{z_a}}^{3} \left( {\frac {{z}^{3}}{{{z_a
}}^{3}}}-{\frac {{z_a}}{z}} \right)}{{L}^{4} \left( {{z_b}}^{4}-
{{z_a}}^{4} \right)}-\frac{10}{3}\,{\frac {{Q}^{2}G\pi \,{\it
z_0}\,{z} ^{3} \left( -{{z_a}}^{2}+{z}^{2} \right) }{{L}^{5}
\left( -{z}^{2}+ {{z_0}}^{2} \right)  \left(
-{{z_a}}^{2}+{{z_0}}^{2} \right) }}$$
$$\psi_b(z)=-\frac{4\,G\pi \,E \left( {\frac {{{z_0}}^{4}}{{{z_a}}^{4}}}-1
 \right) {{z_a}}^{4}{{z_b}}^{3} \left( {\frac {{z}^{3}}{{{z_b
}}^{3}}}-{\frac {{z_b}}{z}} \right)}{{L}^{4} \left( {{z_b}}^{4}-
{{z_a}}^{4} \right)}+\frac{10}{3}\,{\frac {{Q}^{2}G\pi
\,{{z_0}}^{5 } \left( -{{z_b}}^{2}+{z}^{2} \right) }{{L}^{5}z
\left( -{z}^{2}+{{z_0}}^{2} \right)  \left(
-{{z_b}}^{2}+{{z_0}}^{2} \right) }}
$$
and
\begin{eqnarray}
 \Gamma_1&=& {\frac {\arctan \left( R
\left( z_0-z\right) \right)^3 }{\pi }}+ \frac{1}{2}\label{theta(z0-z-m)}\\
  \Gamma_2&=& {\frac {\arctan \left( R \left(z-z_0\right)  \right) ^3}{\pi
}}+\frac{1}{2} \label{theta(z-z0-m)}
\end{eqnarray}
Now one has  to put  conditions (\ref{boundary-d}) on the
regularized functions \be \label{tilde-z}
\frac{z_a}{2L}\left.\frac{d}{dz}\psi_{reg}\right|_{z=\tilde{z}_a}=1\ee
However it is difficult to find $\tilde{z}_a$  from the condition
(\ref{tilde-z}). Instead of finding  $\tilde{z}_a$  from the
condition (\ref{tilde-z}) we propose to use such regularization
that does not change  $z_a$ found from the formal conditions
(\ref{boundary-d}). We can check that the formal $z_a$ in fact
solves also the regularized condition if the regularization is
smooth enough. So, the take $z_a$ and substitute it in the LHS of
the regularized condition (\ref{tilde-z}). We define

$$F_{a,
reg}~\Bigg|_{z=z_a}=\frac{z_a}{2L}\left(\frac{d\psi_a}{dz}\Gamma_1+\frac{d\psi_b}{dz}\Gamma_2\right)~\Bigg|_{z=z_a}\approx1,$$
$$F_{b,reg}~\Bigg|_{z=z_b}=\frac{z_b}{2L}\left(\frac{d\psi_a}{dz}\Gamma_1+\frac{d\psi_b}{dz}\Gamma_2\right)~\Bigg|_{z=z_b}\approx-1.$$

We can calculate  $F_{a, reg}$ and $F_{b, reg}$.  In the following
table we present calculations of $F_{a, reg}$ and $F_{b, reg}$ for
the wide range of parameter of the theory. Results of  calculations at $R=10^4$
are presented in the following  table:

$$ $$
\begin{tabular}{|c|c|c|c|c|c|c|}
  \hline
  $E$,\, $TeV$ & $Q=Q_n/\sqrt{\pi}$,\, $fm^{1/2}$& $z_a$,\, $fm$ & $z_b$,\,$fm$,& $F_a$ & $F_b$ \\
  \hline &&&&&\\
$118.2$ & $40000$ & $0.04928014740$& $4.015208864\cdot10^6$& $0.99997$&$-1.00000$ \\&&&&&\\
  $3$  & $15000$ & $0.08847525298$& $1.019088359\cdot10^6$& $1.00000$&$-1.00000$ \\&&&&&\\
   $0.03$&$40$&$0.7861838575$&$1017.792389$& $1.00000$& $-1.00000$\\&&&&&\\
 \hline
\end{tabular}
$$ $$
Thus, from the table it is evident that   $F_a\approx1,$ $F_b\approx-1.$

\section{Conclusion}\label{Conclusion}
\subsection{Summary}
In this paper we have constructed the phase diagram of the quark gluon
plasma (QGP) formed at a very early stage just after the heavy
ion collision.  In this construction we have used   a holographic dual  model
for the heavy ion collision. In this dual model colliding ions are
described by the charged shock gravitational waves.
Points on the phase diagram correspond to the
QGP or hadronic matter with given temperatures and chemical
potentials. The phase of QGP in dual terms is related to
the case when the collision of shock waves leads to formation of
trapped surface. Hadronic matter and other confined states
correspond to the absence of trapped surface after collision.

 Multiplicity of
the ion collision process  has been  estimated in the dual
language as an area of the trapped surface. We have shown that a non-zero
chemical potential reduces the multiplicity. To plot
 the phase diagram we use two different dual
models of colliding ions. The first model uses the point shock waves
and the second
 the wall shock waves. We have found
  qualitative  agrement of the results.

  A special attention has been devoted to a regularization procedure for calculations performed for
  wall shock waves. On the one hand technically these calculations are
  essential simpler, but on the other hand, this approach, strictly  speaking,
  is incorrect  and requires a regularization. We have shown that a natural regularization does exist.
  Moreover, the proposed regularization does not make calculations to be more complicated
  as compare with naive (direct) calculations. This open new possibility for simple
  calculations for wall shock waves bearing nontrivial matter charges.

\subsection{Further directions}

Head-on collisions of point charged shock waves have only two
parameters. In the dual language they correspond to energy and
chemical potential per nucleus.
 Off-center collisions are also specified
by the impact parameter and the change  of this parameter  can be
associated
 with a dual change from  "non-thermal" peripheral to "thermal" central
 collisions \cite{Shuryak09}.
However, this is still an oversimplification of the problem. The
physics of heavy-ion collision in RHIC is richer and as indicated in
\cite{Shuryak09,Shuryak11}, rapid equilibration and hydrodynamical
behavior experimentally observed at RHIC for collisions of two heavy
ions such as AuAu, does not have the place for deuteron-Au
collisions at the same rapidity. Maybe it is too naive to believe
that the simplest shock wave related by a boost to the Schwarzschild  black hole in AdS
can mimic the nuclear matter in the colliders. However  this simple
shock wave in fact reproduces the interaction of a relativistic
quark with gravity and by this reason, may be considered as a
simplest  candidate to mimic the nuclear matter within the
holographic conjecture. One can try to associate different nuclei
with different
  forms of shock waves. Let us remind in this context that the form of the shock wave
  follows from the  eikonal approximation of the gravity-quark interaction in
  5-dimension \cite{GH,IA-cat}.
  The presence of the electromagnetic field or other fields
  as well as any improvements of the eikonal approximation
  for sure changes the form of the shock waves and it would be interesting to see
  holographical consequences of this consideration.

The obtained lower bound on $N_{\rm charged}$ scales as
$s_{NN}^{1/3}$, which is a faster energy dependence than the
$s_{NN}^{1/4}$ scaling predicted by the Landau model \cite{Landau}
and largely obeyed by the data. If one has a priory a restriction on
allowed energy then one can fit  constants to guaranty that the
experimental data are above the AdS bound. Note that taking into
account the chemical potential permits to increase the allowed
energy. However one cannot expect to much from the chemical
potential corrections.  The relevant chemical potential for baryon
number is not expected to be large, i.e. $\mu_{B}\sim30MeV$ or
$\mu_{B}/T\sim0.15$ for recent experiments at RHIC \cite{rhicdata}
and so any effects will be limited.  However, as has been
mentioned in the text, the  relation
between the value of chemical potential and the value of the
5-dimentional charge is in our disposal and we can assume a huge
ratio of them.

It would be also interesting to try to use plane gravitational waves in AdS$_5$
to describe nonperturbative stages in the gauge theories
and collisions of these waves to describe the QGP formed in the heavy ions collisions.
In the plane case,
the Chandrasekhar-Ferrari-Xanthopoulos duality between colliding
plane gravitational waves and the Kerr black hole solution,
has been used as a model of the BH formation \cite{AVV}.
It would be interesting to generalize this duality  to the AdS case.
This may get  a new insight to a possible dependence of multiplicities  on the rapidity.

\section*{Acknowledgements}

We would  like to thank S.Yu. Vernov and I.V.Volovich   for the
helpful discussions. A.B. is grateful to J. Zaanen, K. Schalm, A.
Parnachev, and M. Cubrovic for illuminating conversations on the
AdS-holography. Our work is partly supported by grant RFFI
11-01-00894-a. The work of A.B. is supported also by Dynasty
foundation and the Dutch Foundation for Fundamental Research on
Matter (FOM).  The work of E.P. is partly supported  by grant
of Russian Ministry of Education and Science 14.740.12.0846.
$$\,$$

\end{document}